\documentclass[12pt,a4]{article}
\usepackage{epsfig}
\usepackage{amsmath}
\usepackage{hhline}
\usepackage{amssymb}
\usepackage{cite}
\usepackage[english]{babel}
\usepackage{graphicx,rotating}

\date{}
\newcommand{\updates}[1]%
        {\fbox{\parbox{\linewidth}{\textbf{Updates from last year:}\\#1}}}

\def\leqsim{\mathbin{\;\raise1pt\hbox{$<$}\kern-8pt\lower3pt\hbox{$\sim$}\;}}
\def\geqsim{\mathbin{\;\raise1pt\hbox{$>$}\kern-8pt\lower3pt\hbox{$\sim$}\;}}

\newcommand{\gev}{\ensuremath{{\rm GeV}}}
\newcommand{\gevc}{\ensuremath{{\rm GeV}/c}}
\newcommand{\gevct}{\ensuremath{{\rm GeV}/c^2}}

\newcommand{\tanb}{\ensuremath{\tan\beta}}

\newcommand{\sLep}{\ensuremath{\tilde{\ell}}}
\newcommand{\sLepL}{\ensuremath{\tilde{\ell}_L}}
\newcommand{\sLepR}{\ensuremath{\tilde{\ell}_R}}

\newcommand{\sEl}{\ensuremath{\tilde{\rm e}}}
\newcommand{\sElL}{\ensuremath{\tilde{\rm e}_L}}
\newcommand{\sElR}{\ensuremath{\tilde{\rm e}_R}}
\newcommand{\sMu}{\ensuremath{\tilde{\mu}}}

\newcommand{\sMuR}{\ensuremath{\tilde{\mu}_R}}
\newcommand{\sTau}{\ensuremath{\tilde{\tau}}}
\newcommand{\sTa}{\ensuremath{\tilde{\tau}}}

\newcommand{\sTauR}{\ensuremath{\tilde{\tau}_R}}
\newcommand{\sTaone}{\ensuremath{\tilde{\tau}_1}}

\newcommand{\Gam}{\ensuremath{\gamma}}

\newcommand{\Chiz}{\ensuremath{{\chi}^{0}}}

\newcommand{\dm}{\ensuremath{\Delta M}}
\newcommand{\mvis}{\ensuremath{M_{\rm vis}}}
\newcommand{\mvisnh}{\ensuremath{M_{\rm vis}({\rm nH})}}
\newcommand{\acop}{\ensuremath{\Phi_{\rm aco}}}
\newcommand{\acopl}{\ensuremath{\Phi_{\rm aco}}}
\newcommand{\acoll}{\ensuremath{\alpha}}
\newcommand{\e}{\ensuremath{E_{12}}}
\newcommand{\eg}{\ensuremath{E_{12}({\rm H})}}

\newcommand{\thetamiss}{\ensuremath{\theta_{\rm miss}}}

\newcommand{\thetamissnh}{\ensuremath{\theta_{\rm miss}({\rm nH})}}
\newcommand{\ptmissnh}{\ensuremath{p_{\rm Tmiss}({\rm nH)}}}
\newcommand{\ptmisstr}{\ensuremath{p_{\rm Tmiss}({\rm tr)}}}
\newcommand{\pmissnh}{\ensuremath{p_{\rm miss}({\rm nH)}}}
\newcommand{\phimiss}{\ensuremath{\phi_{\rm miss}}}
\newcommand{\ptmiss}{\ensuremath{p_{\rm Tmiss}}}
\newcommand{\pmiss}{\ensuremath{p_{\rm miss}}}
\newcommand{\plmiss}{\ensuremath{p_{\rm Lmiss}}}

\newcommand{\gaga}{\ensuremath{\gamma \gamma}}

\def\PLB#1#2#3{{\rm Phys.~Lett.} {\bf{B#1}} (#2) #3}

\def\EPJ#1#2#3{{\rm Eur.~Phys.~J.} {\bf C#1} (#2) #3}

\def\PR#1#2#3{{\rm Phys.~Rep.} {\bf#1} (#2) #3}

\def\NIMA#1#2#3{{\rm Nucl.~Instrum.~and~Methods} {\bf{A#1}} (#2) #3}
\def\CPC#1#2#3{{\rm Comput.~Phys.~Commun.} {\bf#1} (#2) #3}

\newcommand{\pbm}{\ensuremath{{\rm pb}^{-1}}}
\newcommand{\ee}{\ensuremath{{\rm e}^+{\rm e}^-}}

\begin{document}
\font\eightrm=cmr8
\font\ninerm=cmr9

\title{ \null\vspace{3cm}
 Search for scalar leptons in e$^+$e$^-$ collisions \\
 at centre-of-mass energies up to 209~GeV \\
 \vspace{1cm}}
\author{The ALEPH Collaboration$^*)$}
\maketitle

\begin{picture}(160,1)
\put(-10,390){\rm ORGANISATION EUROP\'EENNE POUR LA RECHERCHE NUCL\'EAIRE (CERN)}
\put(60,370){\rm Laboratoire Europ\'een pour la Physique des Particules}
\put(285,280){\parbox[t]{45mm}{\tt CERN-EP/2001-086}}
\put(285,265){\parbox[t]{45mm}{\tt 3rd December 2001}}
\end{picture}

\vspace{.2cm}
\begin{abstract}
\vspace{.2cm}
A search for selectron, smuon and stau pair production
is performed with the data collected by the ALEPH 
detector at LEP at centre-of-mass energies up to  
209~GeV.
The numbers of candidate events are 
consistent with the background predicted by the
Standard Model. Final mass limits from ALEPH 
are reported.
\end{abstract}

\vfill
\centerline{\it (Submitted to Physics Letters B) }
\vskip .5cm
\noindent
--------------------------------------------\hfil\break
{\ninerm $^*)$ See next pages for the list of authors}

\eject
 

\pagestyle{empty}
\newpage
\small
%
%
\newlength{\saveparskip}
\newlength{\savetextheight}
\newlength{\savetopmargin}
\newlength{\savetextwidth}
\newlength{\saveoddsidemargin}
\newlength{\savetopsep}
\setlength{\saveparskip}{\parskip}
\setlength{\savetextheight}{\textheight}
\setlength{\savetopmargin}{\topmargin}
\setlength{\savetextwidth}{\textwidth}
\setlength{\saveoddsidemargin}{\oddsidemargin}
\setlength{\savetopsep}{\topsep}
%
%
\setlength{\parskip}{0.0cm}
\setlength{\textheight}{25.0cm}
\setlength{\topmargin}{-1.5cm}
\setlength{\textwidth}{16 cm}
\setlength{\oddsidemargin}{-0.0cm}
\setlength{\topsep}{1mm}
\pretolerance=10000
\centerline{\large\bf The ALEPH Collaboration}
\footnotesize
\vspace{0.5cm}
{\raggedbottom
\begin{sloppypar}
\samepage\noindent
A.~Heister,
S.~Schael
\nopagebreak
\begin{center}
\parbox{15.5cm}{\sl\samepage
Physikalisches Institut das RWTH-Aachen, D-52056 Aachen, Germany}
\end{center}\end{sloppypar}
\vspace{2mm}
\begin{sloppypar}
\noindent
R.~Barate,
R.~Bruneli\`ere,
I.~De~Bonis,
D.~Decamp,
C.~Goy,
S.~Jezequel,
J.-P.~Lees,
F.~Martin,
E.~Merle,
\mbox{M.-N.~Minard},
B.~Pietrzyk,
B.~Trocm\'e
\nopagebreak
\begin{center}
\parbox{15.5cm}{\sl\samepage
Laboratoire de Physique des Particules (LAPP), IN$^{2}$P$^{3}$-CNRS,
F-74019 Annecy-le-Vieux Cedex, France}
\end{center}\end{sloppypar}
\vspace{2mm}
\begin{sloppypar}
\noindent
G.~Boix,
S.~Bravo,
M.P.~Casado,
M.~Chmeissani,
J.M.~Crespo,
E.~Fernandez,
M.~Fernandez-Bosman,
Ll.~Garrido,$^{15}$
E.~Graug\'{e}s,
J.~Lopez,
M.~Martinez,
G.~Merino,
R.~Miquel,$^{31}$
Ll.M.~Mir,$^{31}$
A.~Pacheco,
D.~Paneque,
H.~Ruiz
\nopagebreak
\begin{center}
\parbox{15.5cm}{\sl\samepage
Institut de F\'{i}sica d'Altes Energies, Universitat Aut\`{o}noma
de Barcelona, E-08193 Bellaterra (Barcelona), Spain$^{7}$}
\end{center}\end{sloppypar}
\vspace{2mm}
\begin{sloppypar}
\noindent
A.~Colaleo,
D.~Creanza,
N.~De~Filippis,
M.~de~Palma,
G.~Iaselli,
G.~Maggi,
M.~Maggi,
S.~Nuzzo,
A.~Ranieri,
G.~Raso,$^{24}$
F.~Ruggieri,
G.~Selvaggi,
L.~Silvestris,
P.~Tempesta,
A.~Tricomi,$^{3}$
G.~Zito
\nopagebreak
\begin{center}
\parbox{15.5cm}{\sl\samepage
Dipartimento di Fisica, INFN Sezione di Bari, I-70126 Bari, Italy}
\end{center}\end{sloppypar}
\vspace{2mm}
\begin{sloppypar}
\noindent
X.~Huang,
J.~Lin,
Q. Ouyang,
T.~Wang,
Y.~Xie,
R.~Xu,
S.~Xue,
J.~Zhang,
L.~Zhang,
W.~Zhao
\nopagebreak
\begin{center}
\parbox{15.5cm}{\sl\samepage
Institute of High Energy Physics, Academia Sinica, Beijing, The People's
Republic of China$^{8}$}
\end{center}\end{sloppypar}
\vspace{2mm}
\begin{sloppypar}
\noindent
D.~Abbaneo,
P.~Azzurri,
T.~Barklow,$^{30}$
O.~Buchm\"uller,$^{30}$
M.~Cattaneo,
F.~Cerutti,
B.~Clerbaux,
H.~Drevermann,
R.W.~Forty,
M.~Frank,
F.~Gianotti,
T.C.~Greening,$^{26}$
J.B.~Hansen,
J.~Harvey,
D.E.~Hutchcroft,
P.~Janot,
B.~Jost,
M.~Kado,$^{31}$
P.~Maley,
P.~Mato,
A.~Moutoussi,
F.~Ranjard,
L.~Rolandi,
D.~Schlatter,
G.~Sguazzoni,
W.~Tejessy,
F.~Teubert,
A.~Valassi,
I.~Videau,
J.J.~Ward
\nopagebreak
\begin{center}
\parbox{15.5cm}{\sl\samepage
European Laboratory for Particle Physics (CERN), CH-1211 Geneva 23,
Switzerland}
\end{center}\end{sloppypar}
\vspace{2mm}
\begin{sloppypar}
\noindent
F.~Badaud,
S.~Dessagne,
A.~Falvard,$^{20}$
D.~Fayolle,
P.~Gay,
J.~Jousset,
B.~Michel,
S.~Monteil,
D.~Pallin,
J.M.~Pascolo,
P.~Perret
\nopagebreak
\begin{center}
\parbox{15.5cm}{\sl\samepage
Laboratoire de Physique Corpusculaire, Universit\'e Blaise Pascal,
IN$^{2}$P$^{3}$-CNRS, Clermont-Ferrand, F-63177 Aubi\`{e}re, France}
\end{center}\end{sloppypar}
\vspace{2mm}
\begin{sloppypar}
\noindent
J.D.~Hansen,
J.R.~Hansen,
P.H.~Hansen,
B.S.~Nilsson,
A.~W\"a\"an\"anen
\nopagebreak
\begin{center}
\parbox{15.5cm}{\sl\samepage
Niels Bohr Institute, 2100 Copenhagen, DK-Denmark$^{9}$}
\end{center}\end{sloppypar}
\vspace{2mm}
\begin{sloppypar}
\noindent
A.~Kyriakis,
C.~Markou,
E.~Simopoulou,
A.~Vayaki,
K.~Zachariadou
\nopagebreak
\begin{center}
\parbox{15.5cm}{\sl\samepage
Nuclear Research Center Demokritos (NRCD), GR-15310 Attiki, Greece}
\end{center}\end{sloppypar}
\vspace{2mm}
\begin{sloppypar}
\noindent
A.~Blondel,$^{12}$
\mbox{J.-C.~Brient},
F.~Machefert,
A.~Roug\'{e},
M.~Swynghedauw,
R.~Tanaka
\linebreak
H.~Videau
\nopagebreak
\begin{center}
\parbox{15.5cm}{\sl\samepage
Laboratoire de Physique Nucl\'eaire et des Hautes Energies, Ecole
Polytechnique, IN$^{2}$P$^{3}$-CNRS, \mbox{F-91128} Palaiseau Cedex, France}
\end{center}\end{sloppypar}
\vspace{2mm}
\begin{sloppypar}
\noindent
V.~Ciulli,
E.~Focardi,
G.~Parrini
\nopagebreak
\begin{center}
\parbox{15.5cm}{\sl\samepage
Dipartimento di Fisica, Universit\`a di Firenze, INFN Sezione di Firenze,
I-50125 Firenze, Italy}
\end{center}\end{sloppypar}
\vspace{2mm}
\begin{sloppypar}
\noindent
A.~Antonelli,
M.~Antonelli,
G.~Bencivenni,
G.~Bologna,$^{4}$
F.~Bossi,
P.~Campana,
G.~Capon,
V.~Chiarella,
P.~Laurelli,
G.~Mannocchi,$^{5}$
F.~Murtas,
G.P.~Murtas,
L.~Passalacqua,
M.~Pepe-Altarelli,$^{25}$
P.~Spagnolo
\nopagebreak
\begin{center}
\parbox{15.5cm}{\sl\samepage
Laboratori Nazionali dell'INFN (LNF-INFN), I-00044 Frascati, Italy}
\end{center}\end{sloppypar}
\vspace{2mm}
\begin{sloppypar}
\noindent
J.~Kennedy,
J.G.~Lynch,
P.~Negus,
V.~O'Shea,
D.~Smith,
A.S.~Thompson
\nopagebreak
\begin{center}
\parbox{15.5cm}{\sl\samepage
Department of Physics and Astronomy, University of Glasgow, Glasgow G12
8QQ,United Kingdom$^{10}$}
\end{center}\end{sloppypar}
\vspace{2mm}
\begin{sloppypar}
\noindent
S.~Wasserbaech
\nopagebreak
\begin{center}
\parbox{15.5cm}{\sl\samepage
Department of Physics, Haverford College, Haverford, PA 19041-1392, U.S.A.}
\end{center}\end{sloppypar}
\vspace{2mm}
\begin{sloppypar}
\noindent
R.~Cavanaugh,
S.~Dhamotharan,
C.~Geweniger,
P.~Hanke,
V.~Hepp,
E.E.~Kluge,
G.~Leibenguth,
A.~Putzer,
K.~Tittel,
S.~Werner,$^{19}$
M.~Wunsch$^{19}$
\nopagebreak
\begin{center}
\parbox{15.5cm}{\sl\samepage
Kirchhoff-Institut f\"ur Physik, Universit\"at Heidelberg, D-69120
Heidelberg, Germany$^{16}$}
\end{center}\end{sloppypar}
\vspace{2mm}
\begin{sloppypar}
\noindent
R.~Beuselinck,
D.M.~Binnie,
W.~Cameron,
G.~Davies,
P.J.~Dornan,
M.~Girone,$^{1}$
R.D.~Hill,
N.~Marinelli,
J.~Nowell,
H.~Przysiezniak,$^{2}$
S.A.~Rutherford,
J.K.~Sedgbeer,
J.C.~Thompson,$^{14}$
R.~White
\nopagebreak
\begin{center}
\parbox{15.5cm}{\sl\samepage
Department of Physics, Imperial College, London SW7 2BZ,
United Kingdom$^{10}$}
\end{center}\end{sloppypar}
\vspace{2mm}
\begin{sloppypar}
\noindent
V.M.~Ghete,
P.~Girtler,
E.~Kneringer,
D.~Kuhn,
G.~Rudolph
\nopagebreak
\begin{center}
\parbox{15.5cm}{\sl\samepage
Institut f\"ur Experimentalphysik, Universit\"at Innsbruck, A-6020
Innsbruck, Austria$^{18}$}
\end{center}\end{sloppypar}
\vspace{2mm}
\begin{sloppypar}
\noindent
E.~Bouhova-Thacker,
C.K.~Bowdery,
D.P.~Clarke,
G.~Ellis,
A.J.~Finch,
F.~Foster,
G.~Hughes,
R.W.L.~Jones,
M.R.~Pearson,
N.A.~Robertson,
M.~Smizanska
\nopagebreak
\begin{center}
\parbox{15.5cm}{\sl\samepage
Department of Physics, University of Lancaster, Lancaster LA1 4YB,
United Kingdom$^{10}$}
\end{center}\end{sloppypar}
\vspace{2mm}
\begin{sloppypar}
\noindent
V.~Lemaitre
\nopagebreak
\begin{center}
\parbox{15.5cm}{\sl\samepage
Institut de Physique Nucl\'eaire, D\'epartement de Physique, Universit\'e Catholique de Louvain, 1348 Louvain-la-Neuve, Belgium}
\end{center}\end{sloppypar}
\vspace{2mm}
\begin{sloppypar}
\noindent
U.~Blumenschein,
F.~H\"olldorfer,
K.~Jakobs,
F.~Kayser,
K.~Kleinknecgt,
A.-S.~M\"uller,
G.~Quast,$^{6}$
B.~Renk,
H.-G.~Sander,
S.~Schmeling,
H.~Wachsmuth,
C.~Zeitnitz,
T.~Ziegler
\nopagebreak
\begin{center}
\parbox{15.5cm}{\sl\samepage
Institut f\"ur Physik, Universit\"at Mainz, D-55099 Mainz, Germany$^{16}$}
\end{center}\end{sloppypar}
\vspace{2mm}
\begin{sloppypar}
\noindent
A.~Bonissent,
J.~Carr,
P.~Coyle,
C.~Curtil,
A.~Ealet,
D.~Fouchez,
O.~Leroy,
T.~Kachelhoffer,
P.~Payre,
D.~Rousseau,
A.~Tilquin
\nopagebreak
\begin{center}
\parbox{15.5cm}{\sl\samepage
Centre de Physique des Particules de Marseille, Univ M\'editerran\'ee,
IN$^{2}$P$^{3}$-CNRS, F-13288 Marseille, France}
\end{center}\end{sloppypar}
\vspace{2mm}
\begin{sloppypar}
\noindent
F.~Ragusa
\nopagebreak
\begin{center}
\parbox{15.5cm}{\sl\samepage
Dipartimento di Fisica, Universit\`a di Milano e INFN Sezione di
Milano, I-20133 Milano, Italy.}
\end{center}\end{sloppypar}
\vspace{2mm}
\begin{sloppypar}
\noindent
A.~David,
H.~Dietl,
G.~Ganis,$^{27}$
K.~H\"uttmann,
G.~L\"utjens,
C.~Mannert,
W.~M\"anner,
\mbox{H.-G.~Moser},
R.~Settles,
H.~Stenzel,
G.~Wolf
\nopagebreak
\begin{center}
\parbox{15.5cm}{\sl\samepage
Max-Planck-Institut f\"ur Physik, Werner-Heisenberg-Institut,
D-80805 M\"unchen, Germany\footnotemark[16]}
\end{center}\end{sloppypar}
\vspace{2mm}
\begin{sloppypar}
\noindent
J.~Boucrot,
O.~Callot,
M.~Davier,
L.~Duflot,
\mbox{J.-F.~Grivaz},
Ph.~Heusse,
A.~Jacholkowska,$^{20}$
C.~Loomis,
L.~Serin,
\mbox{J.-J.~Veillet},
J.-B.~de~Vivie~de~R\'egie,$^{28}$
C.~Yuan
\nopagebreak
\begin{center}
\parbox{15.5cm}{\sl\samepage
Laboratoire de l'Acc\'el\'erateur Lin\'eaire, Universit\'e de Paris-Sud,
IN$^{2}$P$^{3}$-CNRS, F-91898 Orsay Cedex, France}
\end{center}\end{sloppypar}
\vspace{2mm}
\begin{sloppypar}
\noindent
G.~Bagliesi,
T.~Boccali,
L.~Fo\`a,
A.~Giammanco,
A.~Giassi,
F.~Ligabue,
A.~Messineo,
F.~Palla,
G.~Sanguinetti,
A.~Sciab\`a,
R.~Tenchini,$^{1}$
A.~Venturi,$^{1}$
P.G.~Verdini
\samepage
\begin{center}
\parbox{15.5cm}{\sl\samepage
Dipartimento di Fisica dell'Universit\`a, INFN Sezione di Pisa,
e Scuola Normale Superiore, I-56010 Pisa, Italy}
\end{center}\end{sloppypar}
\vspace{2mm}
\begin{sloppypar}
\noindent
O.~Awunor,
G.A.~Blair,
J.~Coles,
G.~Cowan,
A.~Garcia-Bellido,
M.G.~Green,
L.T.~Jones,
T.~Medcalf,
A.~Misiejuk,
J.A.~Strong,
P.~Teixeira-Dias
\nopagebreak
\begin{center}
\parbox{15.5cm}{\sl\samepage
Department of Physics, Royal Holloway \& Bedford New College,
University of London, Egham, Surrey TW20 OEX, United Kingdom$^{10}$}
\end{center}\end{sloppypar}
\vspace{2mm}
\begin{sloppypar}
\noindent
R.W.~Clifft,
T.R.~Edgecock,
P.R.~Norton,
I.R.~Tomalin
\nopagebreak
\begin{center}
\parbox{15.5cm}{\sl\samepage
Particle Physics Dept., Rutherford Appleton Laboratory,
Chilton, Didcot, Oxon OX11 OQX, United Kingdom$^{10}$}
\end{center}\end{sloppypar}
\vspace{2mm}
\begin{sloppypar}
\noindent
\mbox{B.~Bloch-Devaux},$^{1}$
D.~Boumediene,
P.~Colas,
B.~Fabbro,
E.~Lan\c{c}on,
\mbox{M.-C.~Lemaire},
E.~Locci,
P.~Perez,
J.~Rander,
\mbox{J.-F.~Renardy},
A.~Rosowsky,
P.~Seager,$^{13}$
A.~Trabelsi,$^{21}$
B.~Tuchming,
B.~Vallage
\nopagebreak
\begin{center}
\parbox{15.5cm}{\sl\samepage
CEA, DAPNIA/Service de Physique des Particules,
CE-Saclay, F-91191 Gif-sur-Yvette Cedex, France$^{17}$}
\end{center}\end{sloppypar}
\vspace{2mm}
\begin{sloppypar}
\noindent
N.~Konstantinidis,
A.M.~Litke,
G.~Taylor
\nopagebreak
\begin{center}
\parbox{15.5cm}{\sl\samepage
Institute for Particle Physics, University of California at
Santa Cruz, Santa Cruz, CA 95064, USA$^{22}$}
\end{center}\end{sloppypar}
\vspace{2mm}
\begin{sloppypar}
\noindent
C.N.~Booth,
S.~Cartwright,
F.~Combley,$^{4}$
P.N.~Hodgson,
M.~Lehto,
L.F.~Thompson
\nopagebreak
\begin{center}
\parbox{15.5cm}{\sl\samepage
Department of Physics, University of Sheffield, Sheffield S3 7RH,
United Kingdom$^{10}$}
\end{center}\end{sloppypar}
\vspace{2mm}
\begin{sloppypar}
\noindent
K.~Affholderbach,$^{23}$
A.~B\"ohrer,
S.~Brandt,
C.~Grupen,
J.~Hess,
A.~Ngac,
G.~Prange,
U.~Sieler
\nopagebreak
\begin{center}
\parbox{15.5cm}{\sl\samepage
Fachbereich Physik, Universit\"at Siegen, D-57068 Siegen, Germany$^{16}$}
\end{center}\end{sloppypar}
\vspace{2mm}
\begin{sloppypar}
\noindent
C.~Borean,
G.~Giannini
\nopagebreak
\begin{center}
\parbox{15.5cm}{\sl\samepage
Dipartimento di Fisica, Universit\`a di Trieste e INFN Sezione di Trieste,
I-34127 Trieste, Italy}
\end{center}\end{sloppypar}
\vspace{2mm}
\begin{sloppypar}
\noindent
H.~He,
J.~Putz,
J.~Rothberg
\nopagebreak
\begin{center}
\parbox{15.5cm}{\sl\samepage
Experimental Elementary Particle Physics, University of Washington, Seattle,
WA 98195 U.S.A.}
\end{center}\end{sloppypar}
\vspace{2mm}
\begin{sloppypar}
\noindent
S.R.~Armstrong,
K.~Berkelman,
K.~Cranmer,
D.P.S.~Ferguson,
Y.~Gao,$^{29}$
S.~Gonz\'{a}lez,
O.J.~Hayes,
H.~Hu,
S.~Jin,
J.~Kile,
P.A.~McNamara III,
J.~Nielsen,
Y.B.~Pan,
\mbox{J.H.~von~Wimmersperg-Toeller}, 
W.~Wiedenmann,
J.~Wu,
Sau~Lan~Wu,
X.~Wu,
G.~Zobernig
\nopagebreak
\begin{center}
\parbox{15.5cm}{\sl\samepage
Department of Physics, University of Wisconsin, Madison, WI 53706,
USA$^{11}$}
\end{center}\end{sloppypar}
\vspace{2mm}
\begin{sloppypar}
\noindent
G.~Dissertori
\nopagebreak
\begin{center}
\parbox{15.5cm}{\sl\samepage
Institute for Particle Physics, ETH H\"onggerberg, HPK, 8093 Z\"urich,
Switzerland.}
\end{center}\end{sloppypar}
}
\footnotetext[1]{Also at CERN, 1211 Geneva 23, Switzerland.}
\footnotetext[2]{Now at LAPP, 74019 Annecy-le-Vieux, France}
\footnotetext[3]{Also at Dipartimento di Fisica di Catania and INFN Sezione di
 Catania, 95129 Catania, Italy.}
\footnotetext[4]{Deceased.}
\footnotetext[5]{Also Istituto di Cosmo-Geofisica del C.N.R., Torino,
Italy.}
\footnotetext[6]{Now at Institut f\"ur Experimentelle Kernphysik, Universit\"at Karlsruhe, 76128 Karlsruhe, Germany.}
\footnotetext[7]{Supported by CICYT, Spain.}
\footnotetext[8]{Supported by the National Science Foundation of China.}
\footnotetext[9]{Supported by the Danish Natural Science Research Council.}
\footnotetext[10]{Supported by the UK Particle Physics and Astronomy Research
Council.}
\footnotetext[11]{Supported by the US Department of Energy, grant
DE-FG0295-ER40896.}
\footnotetext[12]{Now at Departement de Physique Corpusculaire, Universit\'e de
Gen\`eve, 1211 Gen\`eve 4, Switzerland.}
\footnotetext[13]{Supported by the Commission of the European Communities,
contract ERBFMBICT982874.}
\footnotetext[14]{Also at Rutherford Appleton Laboratory, Chilton, Didcot, UK.}
\footnotetext[15]{Permanent address: Universitat de Barcelona, 08208 Barcelona,
Spain.}
\footnotetext[16]{Supported by the Bundesministerium f\"ur Bildung,
Wissenschaft, Forschung und Technologie, Germany.}
\footnotetext[17]{Supported by the Direction des Sciences de la
Mati\`ere, C.E.A.}
\footnotetext[18]{Supported by the Austrian Ministry for Science and Transport.}
\footnotetext[19]{Now at SAP AG, 69185 Walldorf, Germany}
\footnotetext[20]{Now at Groupe d' Astroparticules de Montpellier, Universit\'e de Montpellier II, 34095 Montpellier, France.}
\footnotetext[21]{Now at D\'epartement de Physique, Facult\'e des Sciences de Tunis, 1060 Le Belv\'ed\`ere, Tunisia.}
\footnotetext[22]{Supported by the US Department of Energy,
grant DE-FG03-92ER40689.}
\footnotetext[23]{Now at Skyguide, Swissair Navigation Services, Geneva, Switzerland.}
\footnotetext[24]{Also at Dipartimento di Fisica e Tecnologie Relative, Universit\`a di Palermo, Palermo, Italy.}
\footnotetext[25]{Now at CERN, 1211 Geneva 23, Switzerland.}
\footnotetext[26]{Now at Honeywell, Phoenix AZ, U.S.A.}
\footnotetext[27]{Now at INFN Sezione di Roma II, Dipartimento di Fisica, Universit\`a di Roma Tor Vergata, 00133 Roma, Italy.}
\footnotetext[28]{Now at Centre de Physique des Particules de Marseille, Univ M\'editerran\'ee, F-13288 Marseille, France.}
\footnotetext[29]{Also at Department of Physics, Tsinghua University, Beijing, The People's Republic of China.}
\footnotetext[30]{Now at SLAC, Stanford, CA 94309, U.S.A.}
\footnotetext[31]{Now at LBNL, Berkely, CA 94720, U.S.A.}
\setlength{\parskip}{\saveparskip}
\setlength{\textheight}{\savetextheight}
\setlength{\topmargin}{\savetopmargin}
\setlength{\textwidth}{\savetextwidth}
\setlength{\oddsidemargin}{\saveoddsidemargin}
\setlength{\topsep}{\savetopsep}
\normalsize
\newpage
\pagestyle{plain}
\setcounter{page}{1}

\section{Introduction}
\label{sec:intro}

The final results of searches for 
sleptons (\sLep) with the data collected by the ALEPH 
detector at LEP at centre-of-mass energies up to 209\,GeV are
presented in this letter.
These searches are interpreted in the theoretical framework 
of the minimal supersymmetric 
extension of the Standard Model (MSSM)~\cite{mssm}, 
with R-parity conservation and the
assumption that the lightest neutralino is the lightest supersymmetric 
particle (LSP).

Scalar leptons are pair produced at LEP
through $s$-channel exchange of a Z~or a \Gam\
giving rise to either \sLepL \sLepL\ or \sLepR \sLepR.
Selectrons can also be produced through the $t$-channel exchange of a
neutralino, resulting also in \sElL \sElR\ production.
Since the amount of mixing between the two slepton states
(\sLepL\ and \sLepR) is proportional to the ratio of
the lepton mass to the slepton mass, it is negligible for selectrons and
smuons and may be sizeable for staus.
As a consequence, the stau production cross section depends on the
MSSM parameters $A_\tau$, $\mu$ and $\tan\beta$ via stau mixing, while the
smuon production cross section depends only on the smuon mass. A dependence
on $M_2$, $\mu$ and $\tan\beta$ is also present for the selectron cross
section via the couplings with the neutralino in the $t$-channel production.

Sleptons decay predominantly into their 
Standard Model (SM) partners
and the lightest neutralino, $\sLep^{\pm} \rightarrow \ell^{\pm} \Chiz_1$.
The experimental signature is therefore a pair of oppositely-charged,
same-flavour, acoplanar leptons (electrons, muons or taus), accompanied 
with missing energy carried away by the two undetected neutralinos.

Results of previous slepton searches up to $\sqrt{s} = 202$\,\gev\
have been reported by 
ALEPH~\cite{aleph_slep_1997,aleph_slep_1998,aleph_slep_1999}
and by the other LEP collaborations~\cite{lep_slep}.
In this letter the analyses performed on the data taken
in the year 2000 at centre-of-mass energies ranging 
from 204 to 209\,GeV are presented.
The selections are mostly
independent of the centre-of-mass energy except for an
appropriate rescaling of the cuts with $\sqrt{s}$ when relevant.
To derive the final results, all the data collected from 
1997 to 2000 are used, with luminosities and centre-of-mass energies
as shown in Table~\ref{tab:lumi}.

\begin{table}[b]
\begin{center}
\caption{\small Integrated luminosities, centre-of-mass energy
   ranges and mean centre-of-mass energy
   values for data collected by the ALEPH detector from 1997
   to 2000.}
\vspace*{0.2cm}
\begin{tabular}{|c|c|c|c|} 
  \multicolumn{4}{c}{} \\  \hline \hline
\rule{0pt}{4.6mm}
 Year & Luminosity (\pbm) & Energy range (GeV) & $\langle \sqrt{s} \rangle$ (GeV) \\ \hline
 2000 &    9.4            & $207-209$        & 208.0 \\
      &  122.6            & $206-207$        & 206.6 \\
      &   75.3            & $204-206$        & 205.2 \\ \hline
 1999 &   42.0            &   $-$              & 201.6 \\
      &   86.2            &   $-$              & 199.5 \\
      &   79.8            &   $-$              & 195.5 \\
      &   28.9            &   $-$              & 191.6 \\ \hline
 1998 &  173.6            &   $-$              & 188.6 \\ \hline
 1997 &   56.8            &   $-$              & 182.7 \\ 
\hline
\hline
\end{tabular}
\label{tab:lumi}
\end{center}
\end{table}

This letter is organised as follows.
In Section~\ref{sec:detec}, a brief description of the ALEPH detector
is given. The signal and background simulations are presented in 
Section~\ref{sec:monte}.
In Section~\ref{sec:selec}, the selection criteria are described. 
Systematic uncertainties are discussed in Section~\ref{sec:syst}. 
The results are reported in Section~\ref{sec:candi}.

\section{ALEPH detector}
\label{sec:detec}

The ALEPH detector is described in detail in Ref.~\cite{aleph_det}. An account
of its performance and a description of the standard analysis algorithms 
can be found in Ref.~\cite{aleph_det2}. 

In ALEPH, the trajectories of charged particles are measured with a
silicon vertex detector, a cylindrical drift chamber, and a large
time projection chamber (TPC). These detector components are immersed 
in a 1.5\,T axial magnetic field provided by a superconducting 
solenoidal coil.
Reconstructed charged particle trajectories are
called {\it good tracks} if they are reconstructed with at least four space
points in the TPC, a transverse momentum in excess of 200\,MeV/$c$, a polar
angle with respect to the beam such that $\vert\cos\theta\vert < 0.95$, and
originating from within a cylinder of length 10\,cm and radius 2\,cm coaxial
with the beam and centred at the nominal interaction point.

The electromagnetic calorimeter (ECAL), placed between the TPC and the
coil, is a
highly-segmented, 22-radiation-length-thick sandwich of lead planes and
proportional wire chambers. It consists of a barrel and two
endcaps. It is used to identify electrons and photons by
the characteristic longitudinal and transverse developments of the
associated showers, supplemented for low momentum electrons by the
measurement in the TPC of the specific energy loss 
by ionization. Photon
conversions to ${\rm e}^+{\rm e}^-$ are identified as pairs of
oppositely-charged electrons satisfying stringent conditions on their
distance of closest approach and their invariant mass.
The luminosity monitors (LCAL and SiCAL) extend the ECAL
hermeticity down to 34 mrad from the beam axis.
The hadron calorimeter (HCAL) consists of the iron return
yoke of the magnet instrumented with streamer tubes. It provides 
a measurement of the hadronic energy and, together with the 
external muon chambers, efficient identification of muons by 
their characteristic penetration pattern.

Global event quantities (such as total energy and missing
momentum) are determined with an energy-flow algorithm which combines
all the above measurements into charged particles (electrons, muons and
charged hadrons), photons and neutral hadrons, called {\it energy-flow
particles} in the following. 
Tau identification proceeds by clustering the
energy-flow particles into two jets with the Durham
algorithm~\cite{durham}. A jet is called a tau-jet candidate if
it contains one or three good track(s) (or two if it 
contains an identified electron, to allow for asymmetric 
photon conversions) and if the jet invariant mass is smaller 
than 2\,GeV/$c^2$.

\section{Simulated samples}
\label{sec:monte}

The signal samples were generated using {\tt SUSYGEN}~\cite{SUSYGEN98}, 
with final 
state radiation simulated with the {\tt PHOTOS}~\cite{photos} package 
and tau decays simulated with {\tt TAUOLA}~\cite{tauola}.
The slepton masses were varied in steps of 5\,GeV/$c^2$\,(1\,GeV/$c^2$) 
for a mass difference $\Delta M$ with the lightest neutralino 
larger\,(smaller) than 10\,GeV/$c^2$; the mass of the
neutralino was varied steps of 5\,GeV/$c^2$.
Event samples of all SM background processes relevant for the slepton 
search were also generated, {\it i.e.}, dilepton production, 
$\gamma\gamma$ interactions with leptonic
final states, W and Z pair production, and Zee and We$\nu$ production.
The Bhabha process was 
simulated with {\tt BHWIDE}~\cite{bhwide}, muon and tau pair production
with {\tt KORALZ}~\cite{koralz}, the \gaga\ processes 
with {\tt PHOT02}~\cite{phot02},
WW production with {\tt KORALW}~\cite{koralw} and the remaining
four-fermion processes with {\tt PYTHIA}~\cite{pythia}.
A detailed GEANT~\cite{geant} simulation of the detector response 
was applied to both background and signal events.

\section{Description of the selections}
\label{sec:selec}

The final state topology depends on the slepton flavour.
Selectrons and smuons are selected by requiring exactly two good
tracks identified as electrons or muons and with opposite electric charge,
while staus are selected by requiring two to six good tracks, clustered into
two tau-jet candidates.
The final state topology also depends on $\Delta M$. Because the
backgrounds are different for large values of $\Delta M$ (mostly W pair
production) and small values of $\Delta M$ (mostly $\gamma\gamma$
interactions), the selections were optimized for two $\Delta M$ ranges. They
are called ``small-$\Delta M$" and ``large-$\Delta M$" analyses in the
following. The selections were optimized 
by minimizing the signal cross section
expected to be excluded, on average, in absence of 
signal (the $\bar{N}_{95}$ prescription~\cite{nbar95}).
The selection criteria are summarized in Table~\ref{tab:sel_smu}
for selectrons and smuons and in Table~\ref{tab:sta} for staus.
The variables used in both the small- and large-\dm\ analyses are defined
below.

\begin{table}[b]
\begin{center}
\caption{\small Selection criteria for the searches for acoplanar selectrons
 and smuons. The energies are expressed in\,GeV, the momenta in\,\gevc\
 and the masses in\,\gevct. 
 Di-fermion final state events, four-fermion final
 state events and photon conversions are denoted by 
 2f, 4f and $\gamma$ conv., respectively.
 The sliding energy cut is 
 explained in Section~\protect\ref{sec:candi}.\vspace*{0.5cm} }
\vspace*{0.2cm}
\begin{tabular}{|c|c|c|c|} \hline \hline
 \multicolumn{4}{|c|}{Selectron and smuon selection criteria} \\ \hline
 \multicolumn{2}{|c|}{} & \multicolumn{1}{c}{Large $\dm$}
 & \multicolumn{1}{|c|}{Small $\dm$} \\ \cline{2-3}
\hline
\hline
Preselection  &  Good tracks  & \multicolumn{2}{c|} {Two oppositely-charged,
same-flavour,} \\
  &       & \multicolumn{2}{c|} {identified e or $\mu$} \\
\hline
Anti-2f & Acoplanarity     & \multicolumn{2}{c|} { \acop $<$ 170$^{\circ}$ } \\
\hline
Anti-(2f+$\gamma$) & Neutral veto     & \multicolumn{2}{c|} {Cut applied} \\
\hline
Anti-($\gamma$ conv.) & Acollinearity    & \multicolumn{2}{c|} {$\alpha > 2^{\circ}$} \\
\hline
Anti-\gaga  &  Energy              &  $\e = 0$        &  $\eg = 0$   \\ 
            &  within $12^{\circ}$ &                &            \\ \cline{2-4}
 &  Visible mass & $\mvis\ > 4$  & $\mvisnh > 4$  \\ \cline{2-4}
 & $\rho$      & $\rho > 2$ & $\rho > 1$ \\ \cline{2-4}
 &  Missing    & $\ptmiss >3\% \sqrt{s}$ & $\ptmissnh > 1\% \sqrt{s}$ \\
 &  momentum   & if $\vert\cos\phimiss\vert<0.26$,&$\ptmisstr>1\% \sqrt{s}$ \\
 &             & then $\ptmiss>5\% \sqrt{s}$&$\vert\cos\thetamissnh\vert<0.9$ \\ \cline{2-4}
 & Fisher variable  & $-$ & Cut applied \\ \cline{2-4}
 & Momenta of the   & \multicolumn{2}{|c|} {$p_{\rm T1},p_{\rm T2} > 0.5\% \sqrt{s}$}  \\ \cline{1-1} \cline{3-4}
Anti-4f & leptons   &  $p_1 < 46.5\% \sqrt{s}$ & {$p_{\rm T1} < 10\% \sqrt{s}$} \\ \cline{2-4}
 & Visible mass     &    $-$ & $\mvisnh < 20\% \sqrt{s}$  \\ \cline{2-4}
 & Miss. momentum   &    $-$ & $\pmissnh < 10\% \sqrt{s}$  \\  \cline{2-4}
 & \multicolumn{3}{|c|} { Sliding energy cut }  \\
\hline
\hline
\end{tabular}
\label{tab:sel_smu}
\end{center}
\end{table}


\begin{table}[b]
\begin{center}
\caption{\small Selection criteria for the search for acoplanar staus.
  The energies are expressed in\,GeV, the momenta in\,\gevc\
  and the masses in\,\gevct. 
  Di-fermion final state events, four-fermion final
  state events and photon conversions are denoted by 
  2f, 4f and $\gamma$ conv., respectively.
  The $p_\tau$ variable represents the smaller tau-jet momentum.
  The sliding energy cut is explained in 
  Section~\protect\ref{sec:candi}. \vspace*{0.5cm} }
\begin{tabular}{|c|c|c|c|} \hline \hline
  \multicolumn{4}{|c|}{Stau selection criteria} \\ \hline
  \multicolumn{2}{|c|}{} & \multicolumn{1}{c}{Large $\dm$} & \multicolumn{1}{|c|}{Small $\dm$} \\ \cline{2-3}
\hline
\hline
Pre-  & Good tracks   & \multicolumn{2}{c|} {Two to six tracks
clustered in two $\tau$ jet candidates} \\ \cline{3-4}
selection  &            & $p_\tau \times 206/\sqrt{s} > 1.07$  &  $-$ \\
\hline
Anti-     & Acollinearity & \multicolumn{2}{c|} {$\alpha > 2^{\circ}$} \\
($\gamma$ conv.)  &  &  \multicolumn{2}{|c|}{ }  \\
\hline
Anti-             & Neutral veto     & \multicolumn{2}{c|} {yes} \\
(2f+$\gamma$) &                  & \multicolumn{2}{c|} {} \\
\hline
Anti-2f           & Acoplanarity & \multicolumn{2}{c|}
 {$\rho \times 206/\sqrt{s} > (\acop  - 150^{\circ})/7.34 $} \\  \cline{1-1}
Anti-\gaga & and $\rho$  & \multicolumn{2}{c|} {} \\ \cline{2-4}
 & Energy                     & \multicolumn{2}{c|} {\e\ = 0} \\ 
 & within $12^{\circ}$ & \multicolumn{2}{c|} {} \\ \cline{2-4}
 & Visible mass   & \multicolumn{2}{|c|} { $\mvis\ > 4$ } \\  \cline{2-4}

 & Missing    &  $\ptmiss\times 206/\sqrt{s}> $
              &  $\ptmiss\times 206/\sqrt{s}>$  \\
 & momentum   &  $16.4-\mvis\times 0.27$
              &  $20.6 - 4.5 \times  \rho \times 206 / \sqrt{s}$  \\
 &            &  $\ptmiss \times 206/\sqrt{s}<$ & \\
 &            &  $(\thetamiss-10^{\circ})\times  1.24$ & \\
\hline
Anti-4f & Lept. momenta  & $p_1 \times 206/\sqrt{s} < 20.0$  &
                         $p_1 \times 206/\sqrt{s} < 12.6$ \\  \cline{2-4}
        & Visible mass   & $\mvis\ < 80.0$  & $-$ \\  \cline{2-4}
        & \multicolumn{3}{|c|}{Sliding energy cut} \\
\hline
\hline
\multicolumn{4}{ c }{} \\
\end{tabular}
\label{tab:sta}
\end{center}
\end{table}

\begin{itemize}

\item Acoplanarity \acopl \\
The acoplanarity is defined as the angle between the two lepton 
momenta projected onto the plane transverse to the beam.
An acoplanarity cut is used to reject events from 
e$^+$e$^- \rightarrow \ell^+ \ell^-$ 
or \gaga\ $\rightarrow \ell^+ \ell^-$.

\item The $\rho$ variable  \\
The lepton momenta are projected onto the transverse plane
and the thrust axis is computed from the projected momenta.
The $\rho$ variable is defined as the scalar sum of the transverse
momentum components of all energy-flow particles with respect 
to the thrust axis. A cut on $\rho$ reduces the number of 
background events from the e$^+$e$^- \rightarrow \tau^+ \tau^-$
and $\gaga \rightarrow \tau^+ \tau^-$ processes.

\item Variables \mvis, \pmiss, \thetamiss, \phimiss, \ptmiss\ and \plmiss \\
The total energy and momentum is constructed by adding energies and momenta
of all energy-flow particles of the event, 
allowing a total visible mass $M_{\rm
vis}$ to be computed. The missing momentum \pmiss, identical
in magnitude and opposite in direction to the total momentum, defines
the variables \thetamiss\ and \phimiss\ (polar and azimuthal angles of 
its direction), and \ptmiss\ and \plmiss\ (transverse and longitudinal 
components).
Because the small-$\Delta M$ analysis is more sensitive to accidental
double counting between low energy tracks and calorimeter clusters, 
two alternative methods are used 
to determine the total energy and momentum, {\it (i)} without the neutral
hadrons (variables labelled ``nH"); and {\it (ii)} without the neutral
particles (variables labelled ``tr").
The cuts on the visible mass and on the missing momentum
(value and angle) reject background events from \gaga\ processes.
The cut on \ptmiss\ is tightened when the missing momentum points 
to the vertical boundaries between the two halves of the LCAL.

\item Lepton momenta $p_1$, $p_2$ and $p_{\rm T1}$ \\
The variable $p_1$\,$(p_2$) is the momentum of the more\,(less) 
energetic identified electron or muon, if any. For massive neutralinos,
upper cuts on \mvis\ and $p_1$ are effective at rejecting 
leptonic WW events.

\item Activity at small angle \e\ and \eg  \\
Because $\gamma\gamma$ interactions often produce particles 
strongly boosted forward or backward, the requirement that no 
energy be detected at low angles is effective in suppressing 
this background.
The variable \e\ gives the total energy measured 
within 12$^\circ$ of the beam axis in LCAL and 
SiCAL, and in the first cells 
(closest to the beam pipe) of the ECAL endcap calorimeters.
The variable \eg\ takes into account in addition
the energy collected in the first cells of the 
HCAL endcap calorimeters.
This cut introduces an inefficiency due to beam related 
background (Section~\ref{sec:syst}). 

\item Neutral veto \\
Dilepton events with
hard initial or final state radiation at large angle from incoming and
outgoing leptons give rise to acoplanar lepton pairs, and to potentially
large missing energy if the radiated photons go into poorly instrumented
regions of the detector ({\it e.g.}, ECAL module boundaries, overlap between
barrel and endcaps). To reject these events, it is required that the
invariant mass between any good track and any neutral energy-flow
particle, outside a cone of half-angle 10$^\circ$ around the good track and
with an energy in excess of 4\,GeV, be smaller than 2\,GeV/$c^2$.

\item Acollinearity \acoll \\
Single-photon events from e$^+$e$^-$ $\rightarrow \nu \bar{\nu} \gamma$
with a hard radiated photon, followed by a photon 
conversion into an e$^+$e$^-$ pair, are not
rejected by the previous veto, but are characterized by a very small opening
angle between the two tracks (called the acollinearity angle). An acollinearity
cut is therefore effective at rejecting these events.

\item Fisher variable \\
In order to further reduce the large \gaga\ background in the case
of the small-\dm\ selectron or smuon analyses, a 
Fisher discriminant~\cite{fisher} 
analysis is used. This method exploits the remaining
modest difference among \gaga\ events and the signal,
taking into account the correlations 
between the variables \mvis, \ptmiss, \plmiss, $\rho$, $p_1$,
$p_2$, and~\acoll.

\end{itemize}

\section{Systematic effects}
\label{sec:syst}

The main systematic uncertainties on the background and signal 
expectations come from the statistics of the simulated samples 
(from 1.5 to 3.5\%). 
A 2\% systematic uncertainty was estimated for the
lepton identification efficiency.
The inefficiency caused by the E$_{12}$ cuts 
is determined from events 
triggered at random beam crossings.
The corrections for E$_{12}$\,(E$_{12}$(H)) used for 
the year 2000 data are $-$9.5\%\,($-$14.6\%) 
at 205.2\,GeV and $-$8.8\%\,($-$13.7\%) at 206.6 and 208.0\,GeV.
The uncertainties on these corrections are less
than 1\%. 

\begin{figure}[tbhp]
\setlength{\unitlength}{1.0cm}
\begin{center}
\begin{picture}(16.0,18.0)
\put(0.0,8.5){\epsfig{file=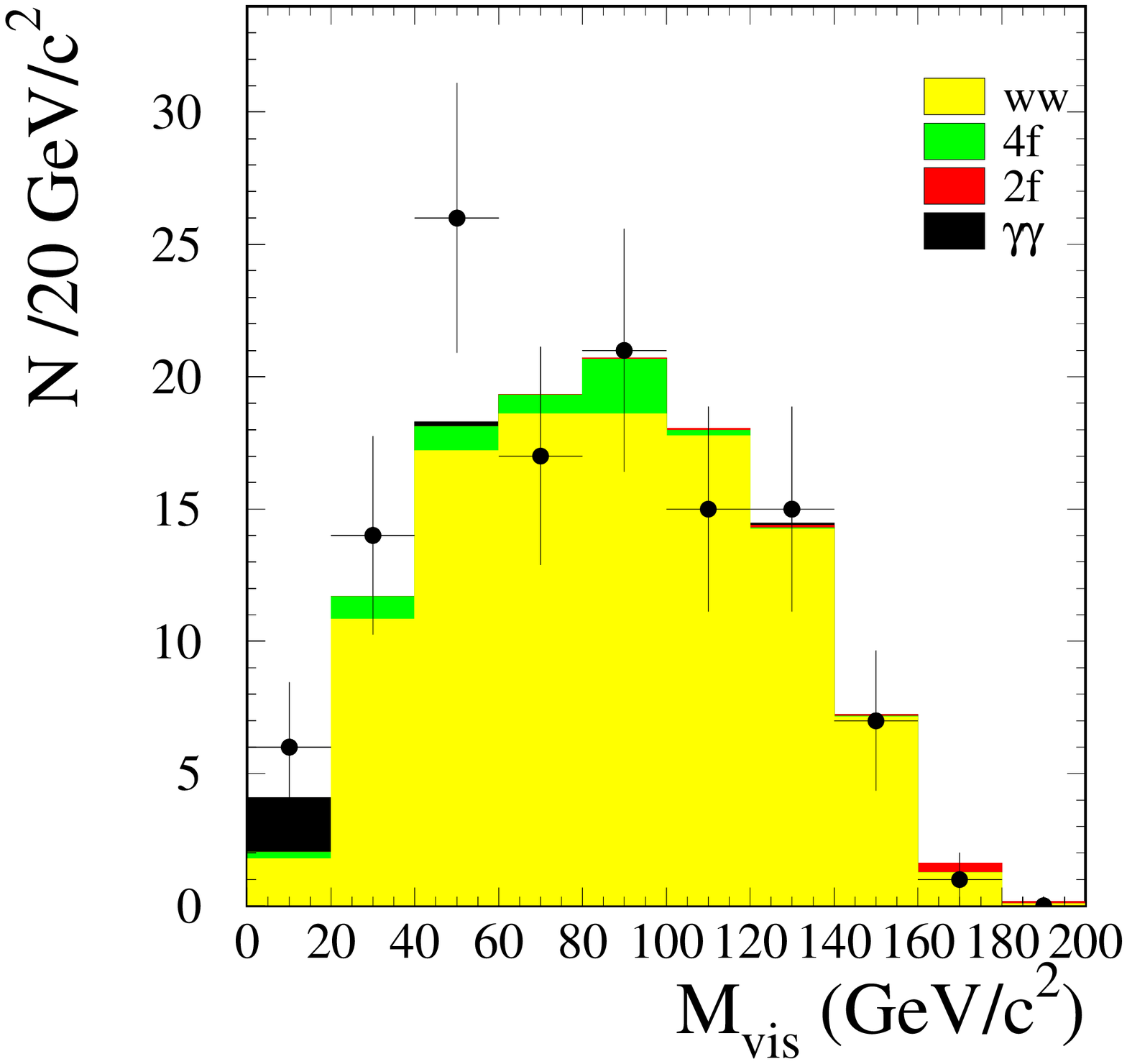,%
            height=8.0cm,width=8.0cm}}
\put(7.5,8.5){\epsfig{file=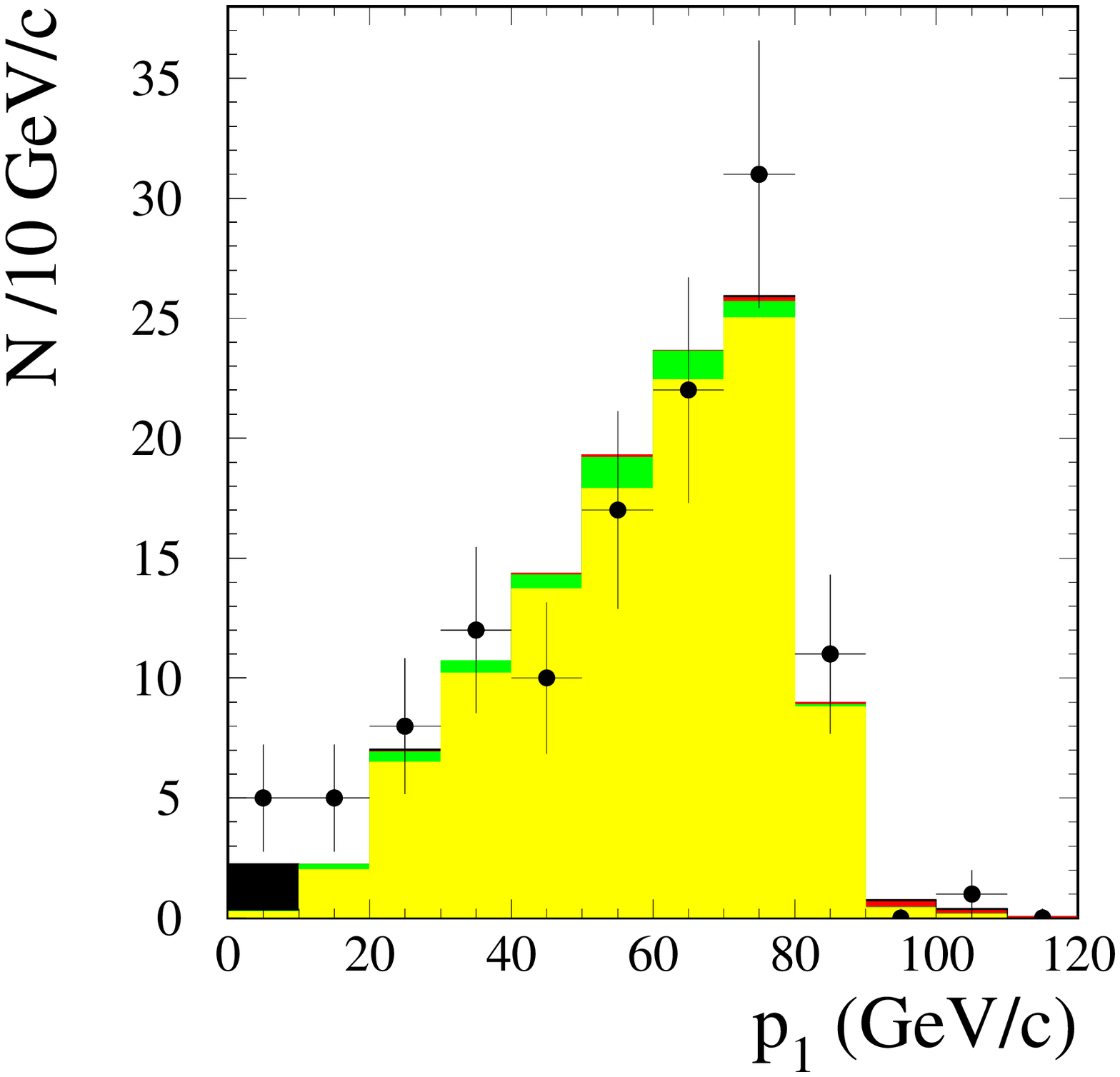,%
            height=8.0cm,width=8.0cm}}
\put(0.0,1.0){\epsfig{file=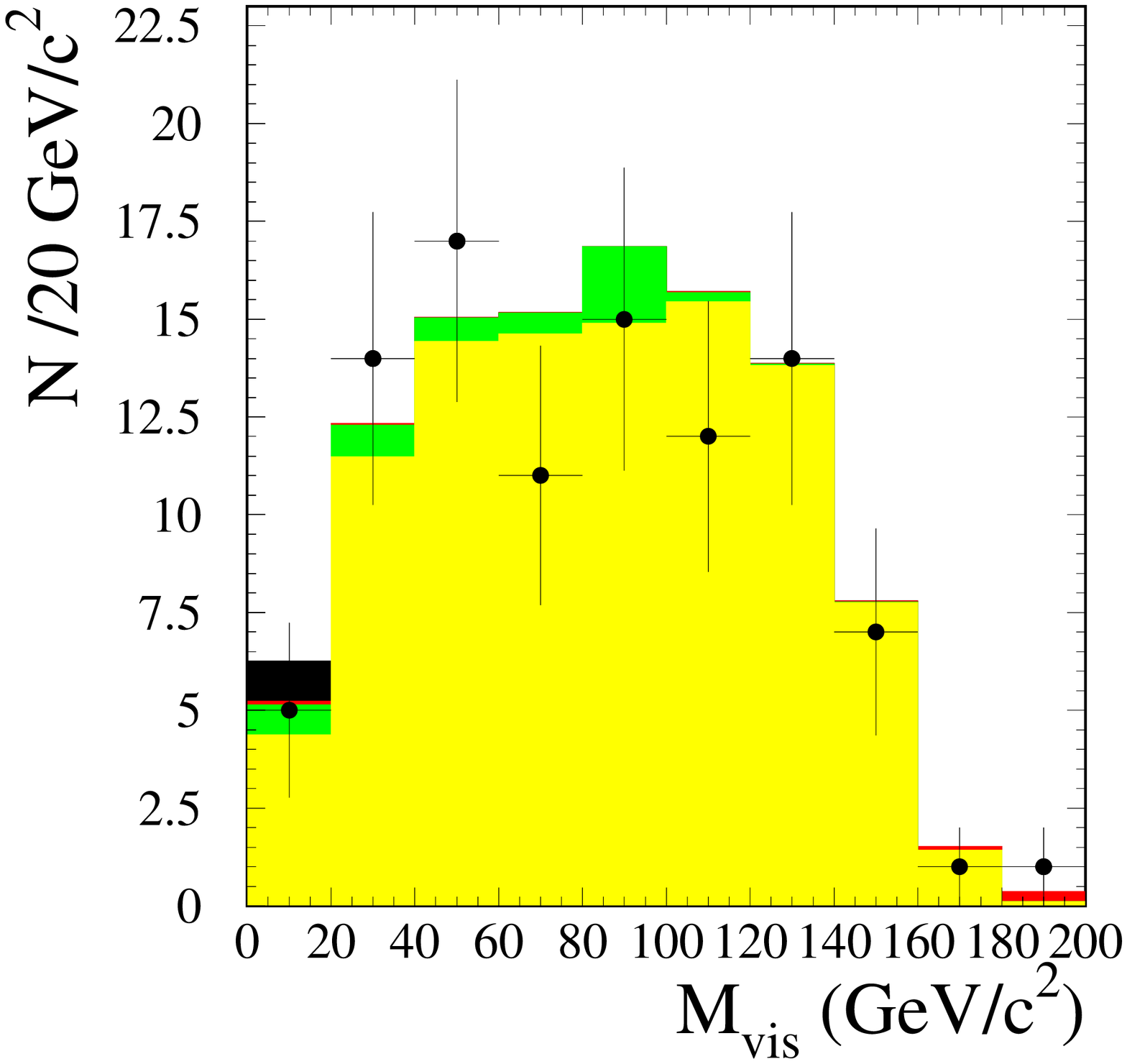,%
            height=8.0cm,width=8.0cm}}
\put(7.5,1.0){\epsfig{file=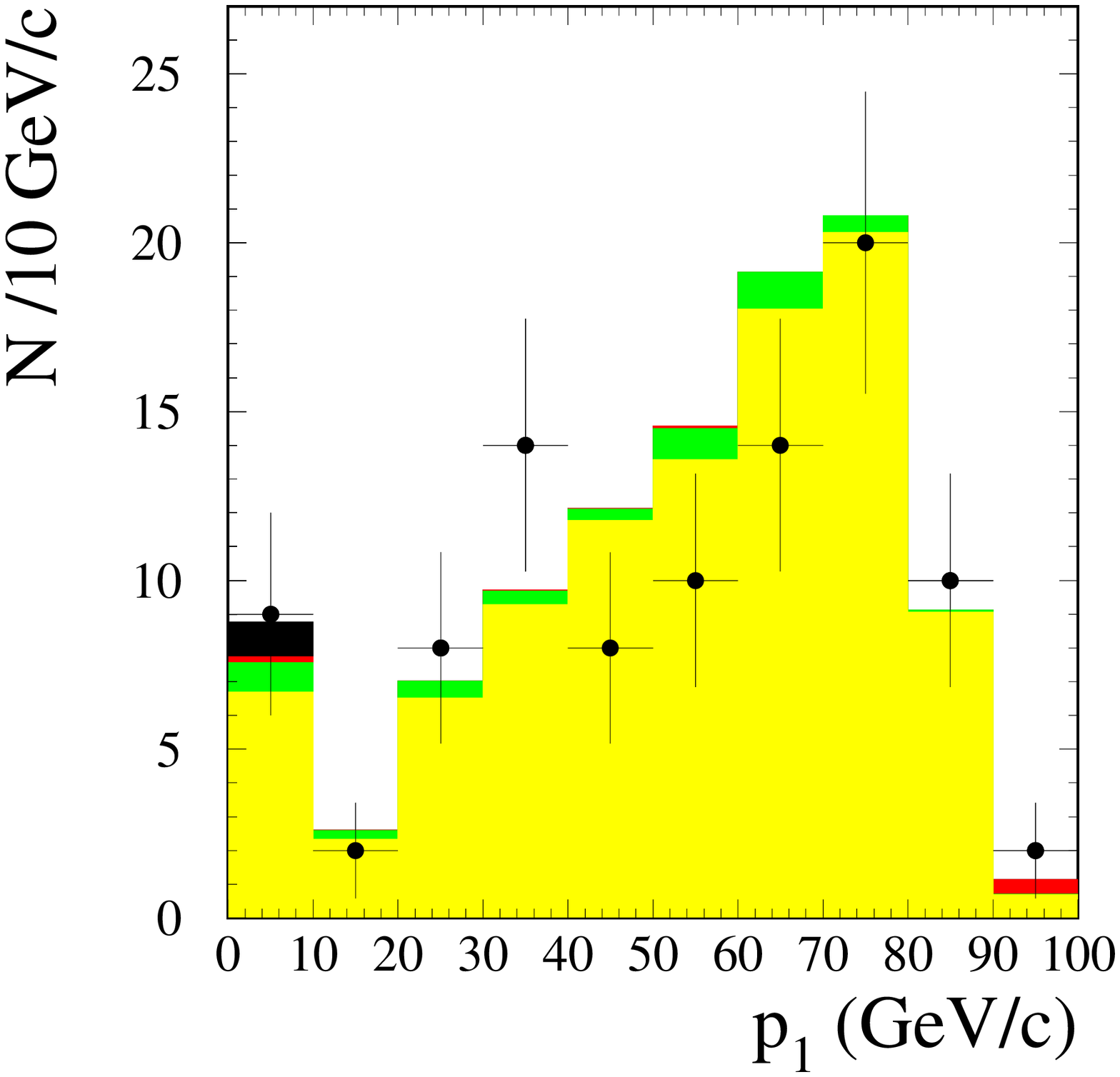,%
            height=8.0cm,width=8.0cm}}
\put(6.8,16.5){\Large \bf A L E P H }
\put(2.0,15.0){(a)}
\put(9.5,15.0){(b)}
\put(6.3,7.5){(c)}
\put(9.5,7.5){(d)}
\end{picture}
\caption{\small Comparison between data (dots with error bars) 
    and expected backgrounds (histograms)
    before anti-four-fermion and lepton identification cuts.
    For selectrons and smuons:
    (a) total visible mass (\mvis) and (b) momentum of the leading 
    lepton ($p_1$). For staus: (c) total visible mass (\mvis) and 
    (d) momentum of the leading lepton identified in the tau 
    jet ($p_1$). The accumulation at zero in (d) corresponds to 
    events with no leptons in either tau jet.} 
\label{fig:contr_pre}
\end{center}
\end{figure}

The numbers of background events and the selection efficiencies
are interpolated from the generated values to the
mean energies of the year 2000 data.
The contribution to the systematic uncertainty
of this interpolation procedure is negligible.
A comparison between data and expected backgrounds
from SM processes is presented in Fig.~\ref{fig:contr_pre}
for the large-\dm\ analysis before anti-four-fermion cuts
and without lepton identification.
At this level, agreement is observed between 
data and the expected backgrounds, dominated by 
W pair production.

\section{Results}
\label{sec:candi}

After the selection criteria described 
in Section~\ref{sec:selec} are applied, 
the numbers of candidate events observed in the data and 
those expected from SM background processes 
are given in Table~\ref{tab:cand_2000_sel_smu} for the selectron and smuon
searches and in Table~\ref{tab:cand_2000_sta} for the stau search.

\begin{table}[t]
\begin{center}
\caption{\small Numbers of candidate events observed in the 
   year 2000 data ($N_{\rm obs}$) 
   and of background events expected from SM processes
   ($N_{\rm bkg}$), in the searches for selectrons and smuons (before the
   sliding energy cut). The numbers in parentheses
   give the WW contribution.}
\begin{tabular}{|c||c|c|c|c||c|c|c|c|}
\multicolumn{9}{c}{} \\  \hline \hline
 $\sqrt{s}$  & \multicolumn{4}{|c||}{\sEl} & \multicolumn{4}{|c|}{\sMu} \\ \cline{2-9}
 (GeV)   & \multicolumn{2}{|c|}{Small \dm} & 
           \multicolumn{2}{|c||}{Large \dm} & 
           \multicolumn{2}{|c|}{Small \dm} & 
           \multicolumn{2}{|c|}{Large \dm} \\  \cline{2-9}
    & $N_{\rm obs}$ & $N_{\rm bkg}$&$N_{\rm obs}$&$N_{\rm bkg}$ 
    & $N_{\rm obs}$ & $N_{\rm bkg}$&$N_{\rm obs}$&$N_{\rm bkg}$\\ \hline
  205.2 & 0 & 0.12 & 15 & 14.0
        & 0 & 0.12 &  8 & 12.5  \\
        &   & (0.09) &    & (12.2)
        &   & (0.07) &    & (11.4) \\ \hline
  206.6 & 0 & 0.19 & 24 & 22.9
        & 0 & 0.20 & 28 & 20.6 \\
        &   & (0.16) &    & (20.0)
        &   & (0.11) &    & (18.8)  \\ \hline
  208.0 & 0 & 0.02 &  0 & 1.76
        & 0 & 0.17 &  3 & 1.59 \\
        &   & (0.01) &    & (1.53)
        &   & (0.08) &    & (1.44)  \\ \hline
\hline
\end{tabular}
\label{tab:cand_2000_sel_smu}
\end{center}
\end{table}
\begin{table}[t]
\begin{center}
\caption{\small Numbers of candidate events observed 
   in the year 2000 data ($N_{\rm obs}$) 
   and of background events expected from SM processes
   ($N_{\rm bkg}$), in the search for staus (before the sliding energy cut).
   The numbers in parentheses give the WW contribution.}
\vspace*{-0.5cm}
\begin{tabular}{|c||c|c|c|c|c|c|} 
        \multicolumn{7}{c}{} \\  
        \multicolumn{7}{c}{} \\  \hline \hline
$\sqrt{s}$  & \multicolumn{6}{|c|}{\sTa}  \\ \cline{2-7}
 (GeV)   & \multicolumn{2}{|c|}{Small \dm} & 
           \multicolumn{2}{|c|}{Large \dm} & 
           \multicolumn{2}{|c|}{Small or Large \dm} \\  \cline{2-7}
         & $N_{\rm obs}$ & $N_{\rm bkg}$ & $N_{\rm obs}$ & $N_{\rm bkg}$ 
         & \ \ \ \ $N_{\rm obs}$ & $N_{\rm bkg}$  \\ \hline
  205.2 &  5 & 6.1  &  1 & 5.4 & 5  & 7.6  \\
        &    & (5.0)   &     & (4.4)   &  & (6.3)  \\ \hline
  206.6 & 12 & 10.1 & 10 & 8.8 & 15 & 12.5  \\
        &    & (8.3)   &     & (7.2)  &  & (10.3)  \\ \hline
  208.0 &  3 & 0.80 &  0    & 0.70 & 3  & 1.01  \\
        &    & (0.68) &     & (0.57) &  & (0.83)  \\ \hline
\hline
\end{tabular}
\vspace*{0.2cm}
\label{tab:cand_2000_sta}
\end{center}
\end{table}

\begin{figure}[hp]
\begin{center}
\setlength{\unitlength}{1.0cm}
\begin{picture}(15.0,20.0)
\put(-0.3,13.6){\epsfig{file=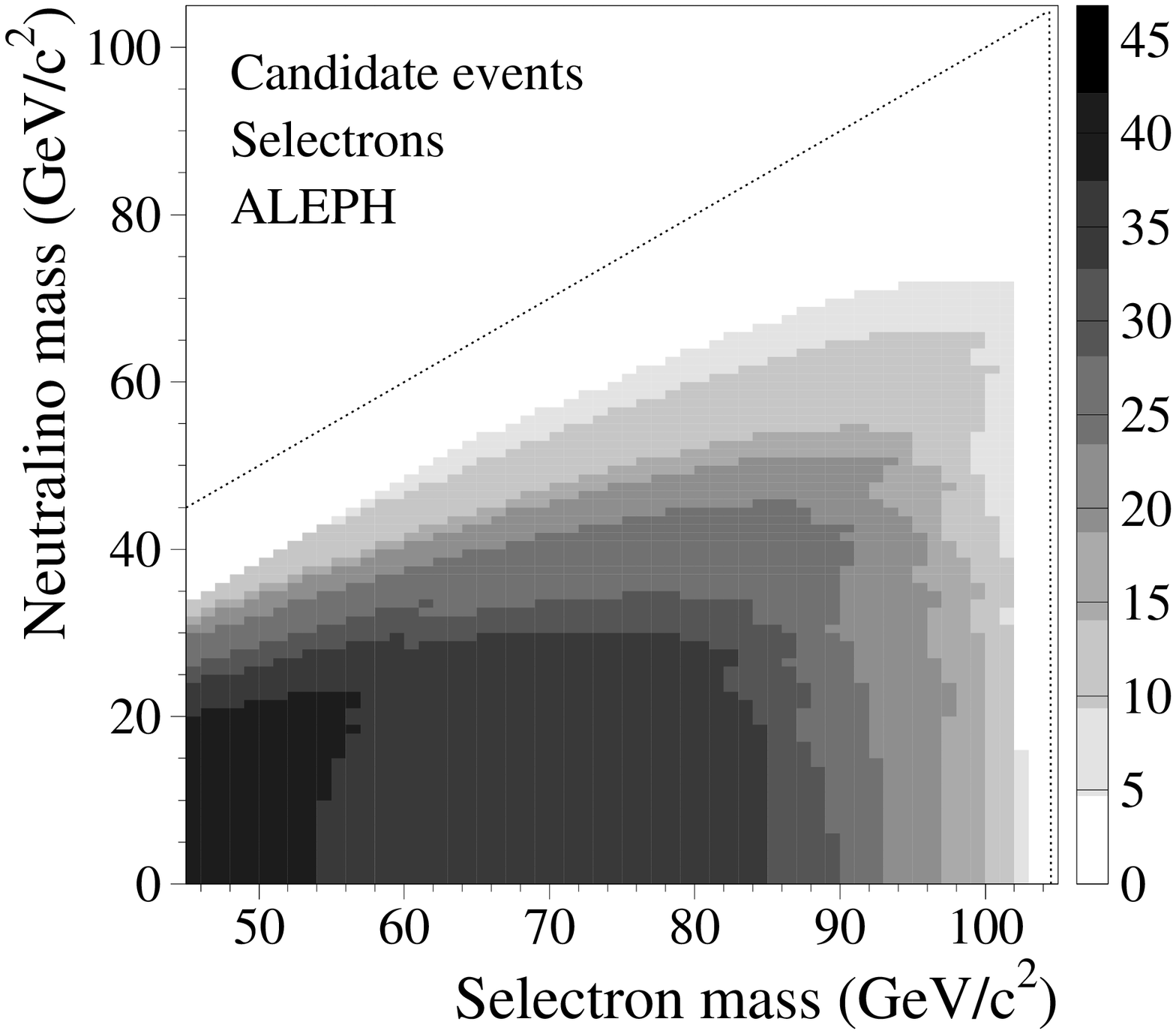,height=7.0cm,width=7.0cm}}
\put(-0.3,6.8){ \epsfig{file=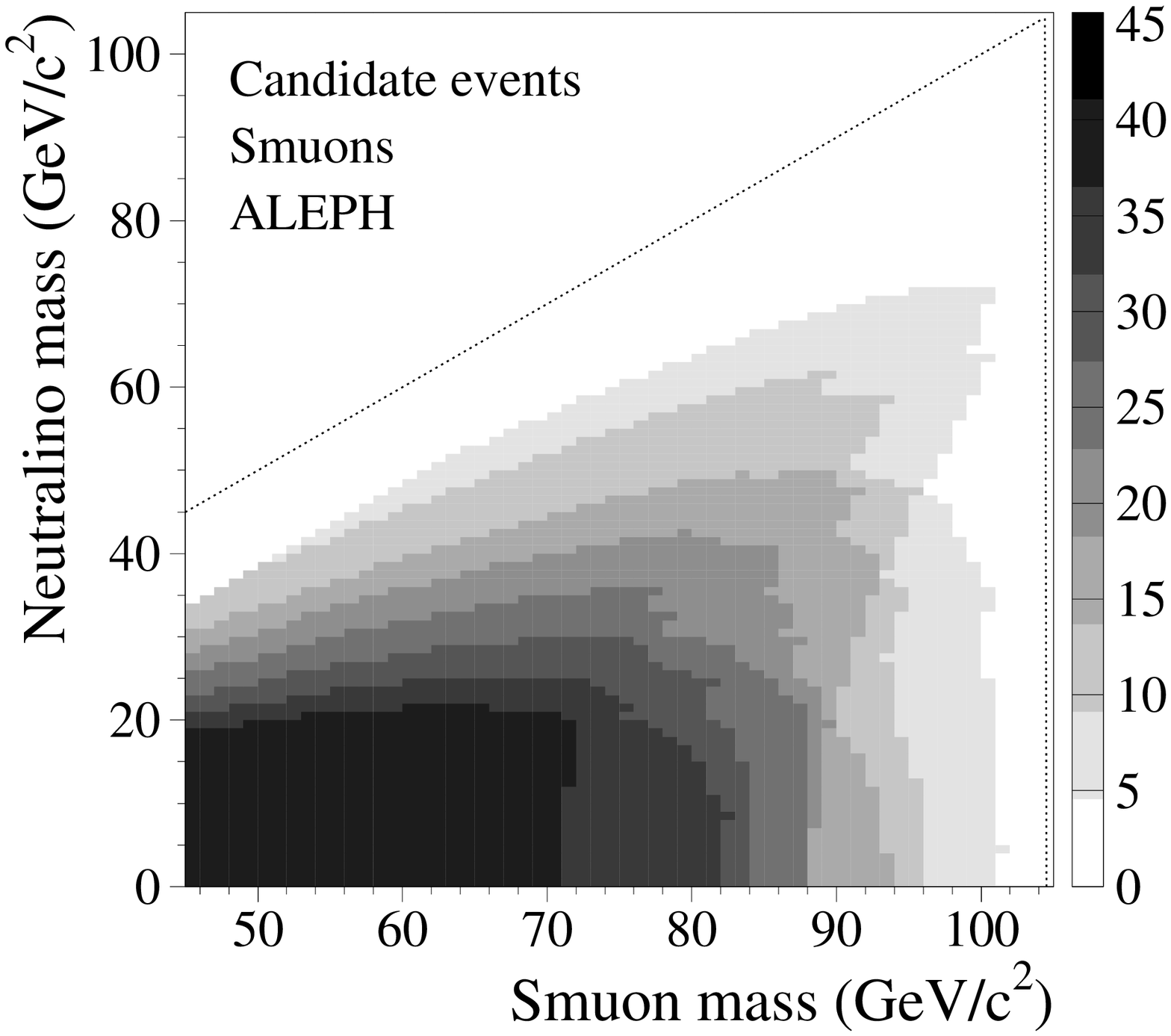,height=7.0cm,width=7.0cm}}
\put(-0.3,0.0){ \epsfig{file=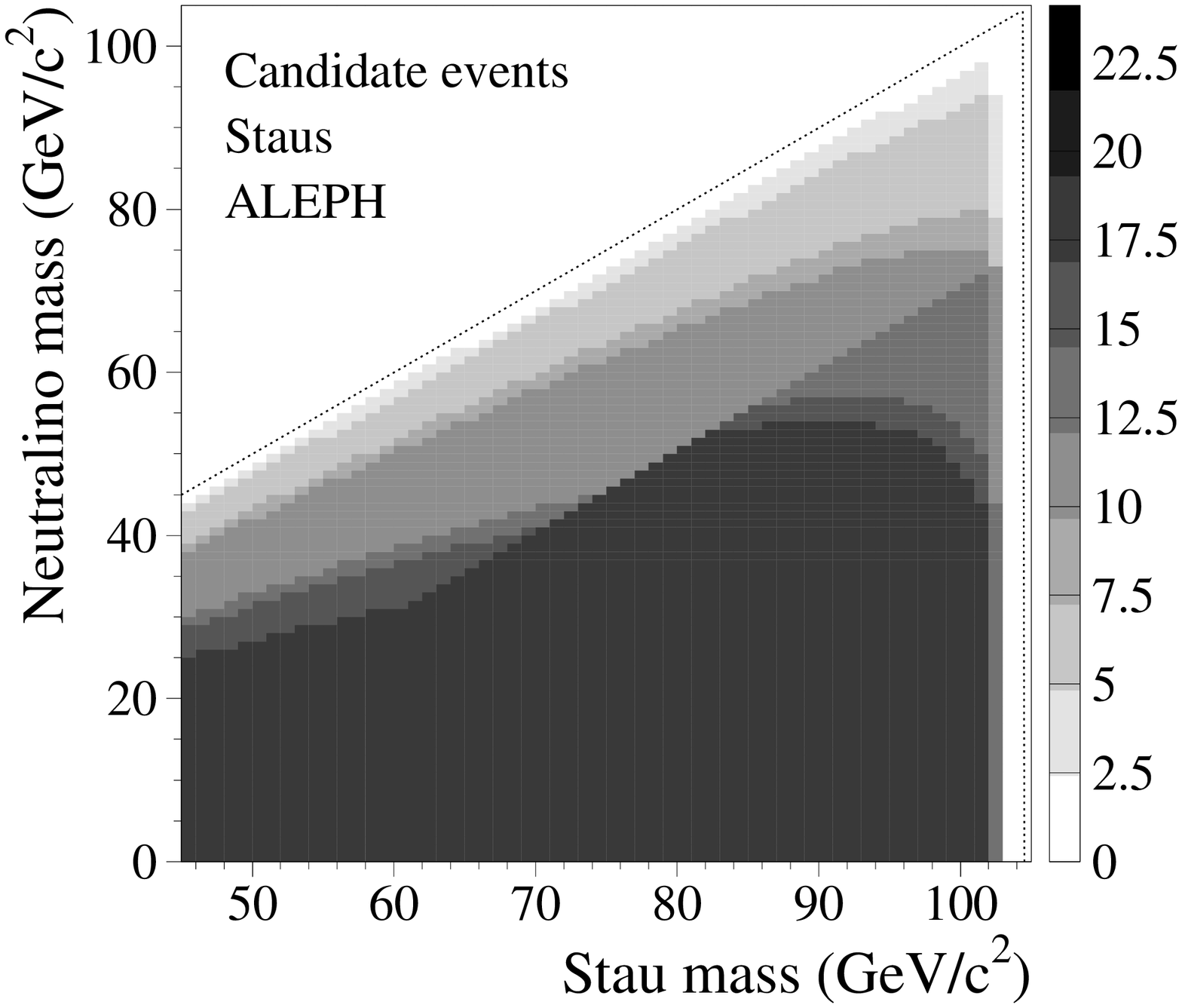,height=7.0cm,width=7.0cm}}
\put(7.5,13.6){ \epsfig{file=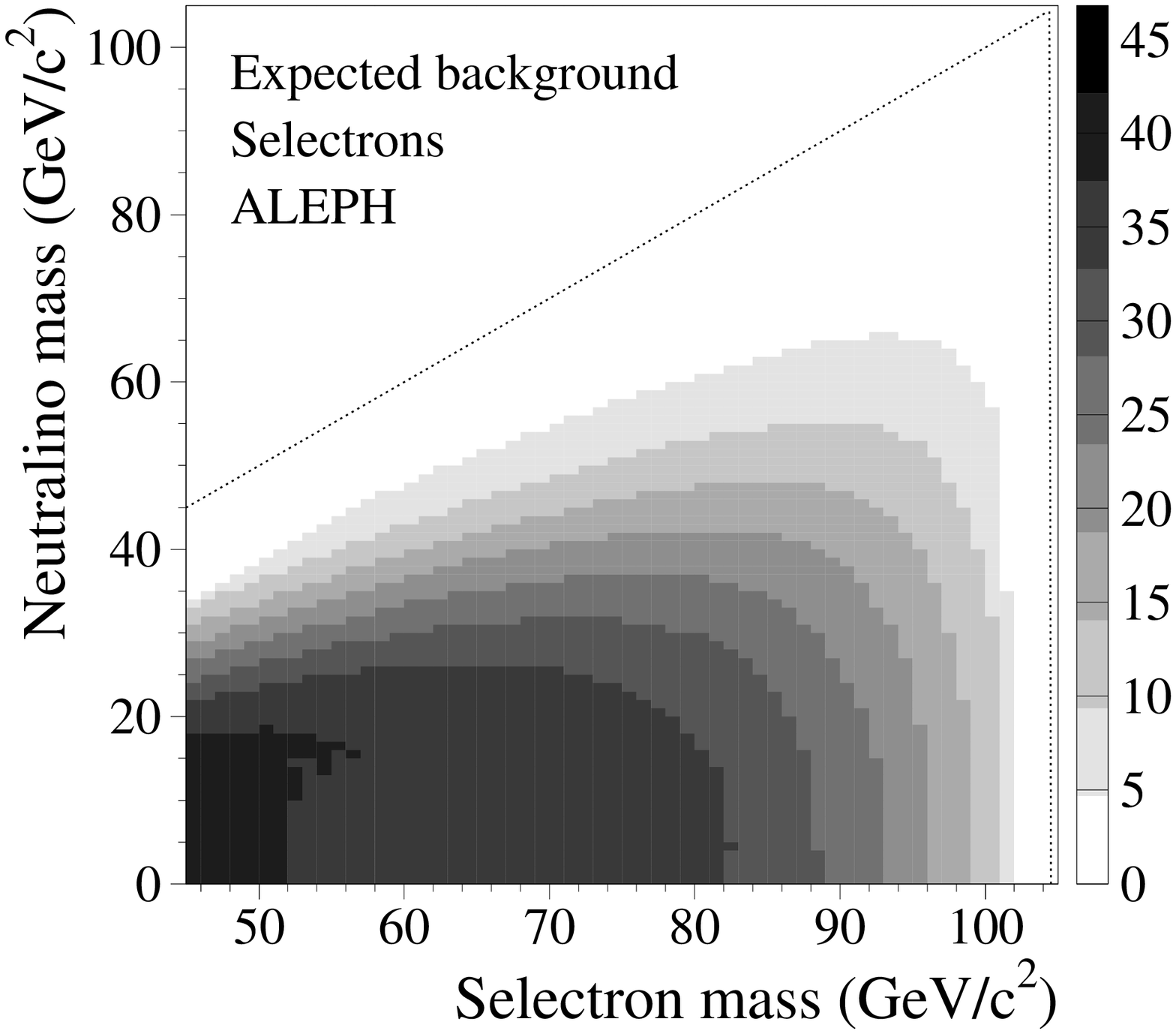,height=7.0cm,width=7.0cm}}
\put(7.5,6.8){  \epsfig{file=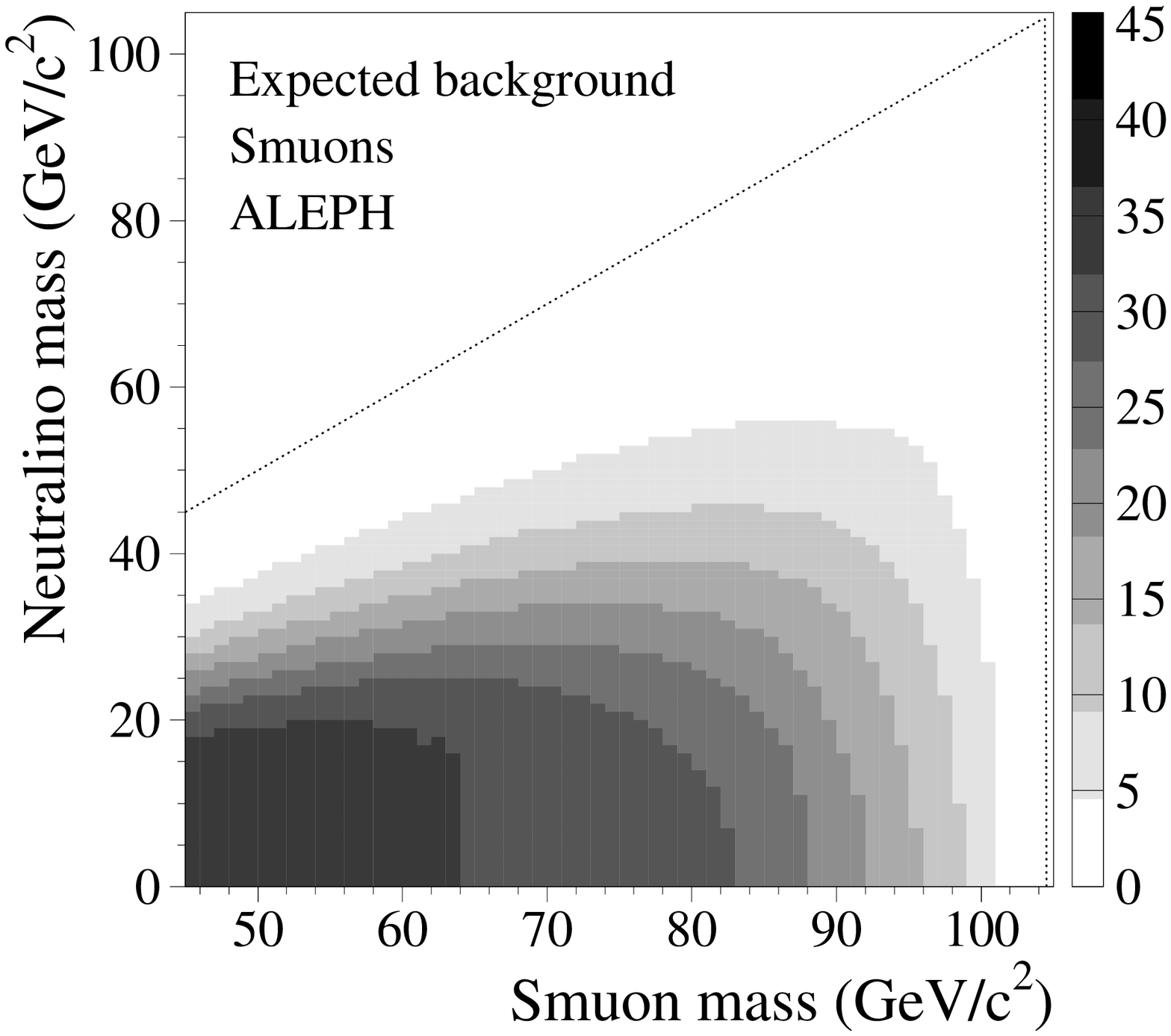,height=7.0cm,width=7.0cm}}
\put(7.5,0.0){  \epsfig{file=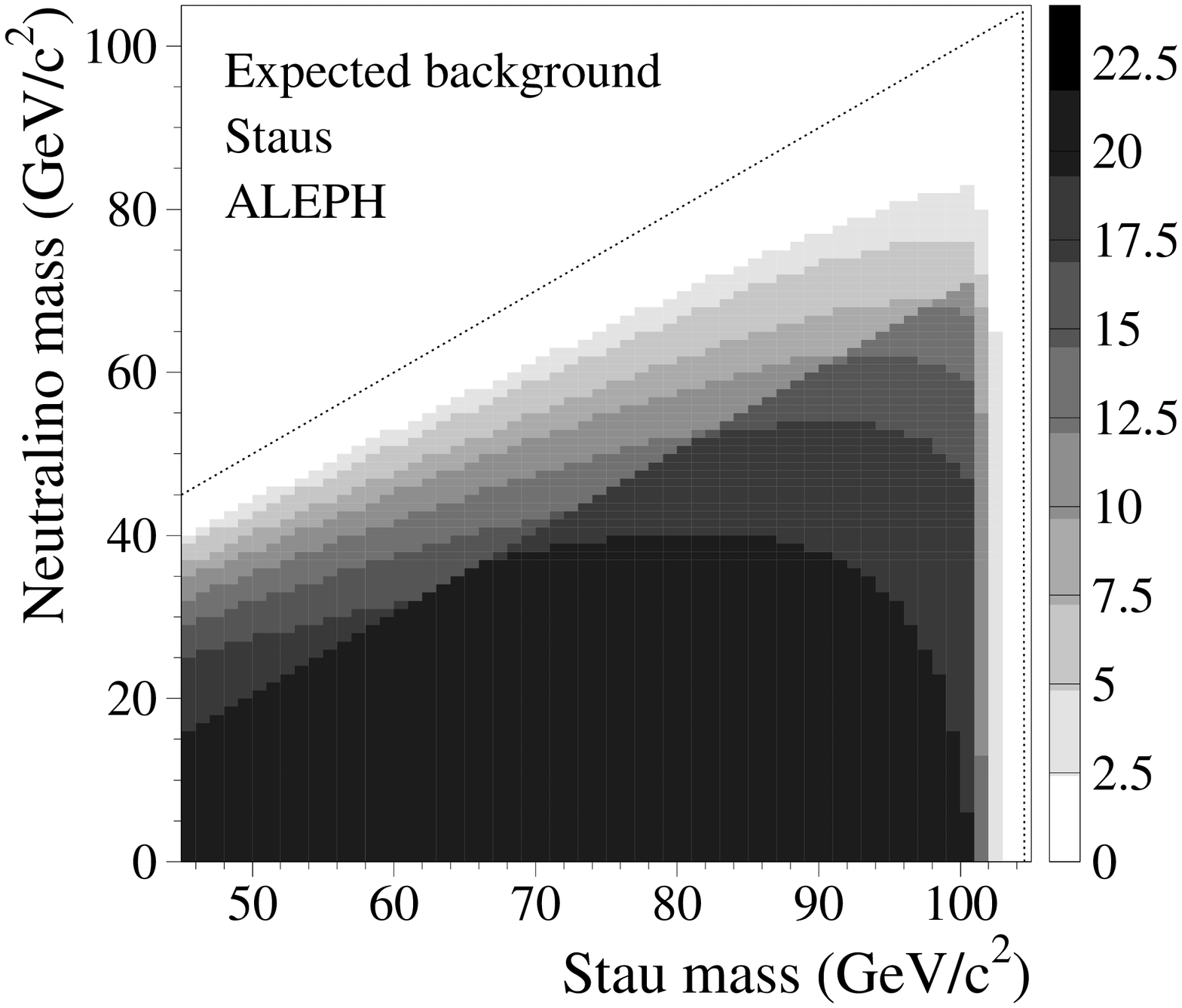,height=7.0cm,width=7.0cm}}
\put(7.0,20.4){(a)}
\put(7.0,13.6){(b)}
\put(7.0,6.9) {(c)}
\end{picture}
\caption{\small Numbers of selected events in the year 2000 data (left) and 
   of expected background events from SM processes (right)
   in the search for (a) selectrons, (b) smuons and (c) staus.
   The density of shading reflects the number of events
   as indicated by the vertical scales.
   The kinematically accessible region at 209\,GeV is indicated by
   the vertical dotted lines. The diagonal dotted lines
   show the boundary of the domain allowed in the MSSM for a 
   neutralino LSP.}
\label{fig:cand_2000}
\end{center}
\end{figure}

The search for sleptons is performed as a function of 
the hypothetical slepton and neutralino masses.
A sliding energy cut is applied (not accounted for 
in Tables~\ref{tab:cand_2000_sel_smu} and~\ref{tab:cand_2000_sta})
which requires that the lepton momenta
be in the range kinematically allowed for a signal with 
the assumed values of slepton and neutralino masses.
For a two-body decay
$\sLep^{\pm} \rightarrow \ell^{\pm} \Chiz_1$,
the minimum and maximum values for the decay lepton energy
are given by
\begin{eqnarray}
 E_\ell^{\rm max,min} =
      \frac{E_{\rm b}}{2} \left( 1 - \frac{m_{\Chiz_1}^2}{m_{\sLep}^2} \right)
      \left[ 1 \pm \sqrt{1 - \frac{m_{\sLep}^2}{E_{\rm b}^2}} \right],
\label{eq.MSSM.sleptons.Elimits}
\end{eqnarray}
where $E_{\rm b}$ is the beam energy and where the lepton mass is neglected. 
In the
case of staus, the exact formula is used to account for the non-negligible
mass of the tau.
An event is compatible with the mass hypothesis
$(m_{\sLep}, m_{\Chiz_1})$ if both decay
leptons satisfy Eq.~(\ref{eq.MSSM.sleptons.Elimits}).
Any given event may therefore be
compatible with a wide mass range, especially for large-\dm\ values.
In the decay of a stau, the resulting tau lepton will itself give rise to
at least one neutrino and the lower limit on the energy
cannot be applied. Only the upper limit can be used, which increases
even further the compatible region.
For the selectron, smuon and stau searches, the numbers of
candidate events and expected background events
are displayed in Fig.~\ref{fig:cand_2000}
as a function of the slepton and the neutralino masses,
after the sliding energy cut is applied.

\subsection{Cross section upper limits}
\label{sec:upper}

The data show no evidence for an excess
of events compared to the estimated background.
The dominant WW background theoretically and experimentally well
under control, is subtracted to derive 95\% C.L. upper limits
on the slepton production cross section times the branching 
ratio into $\ell \Chiz_1$. This partial background subtraction 
is performed with the likelihood ratio method~\cite{Read}.
Data from previous years are combined in the likelihood to increase
the sensitivity. The systematic uncertainties on the efficiencies are
taken into account with the method of Ref.~\cite{Cousins}.

\begin{figure}[p]
\begin{center}
\setlength{\unitlength}{1.0cm}
\begin{picture}(15.0,20.0)
\put(-0.3,14.0){\epsfig{file=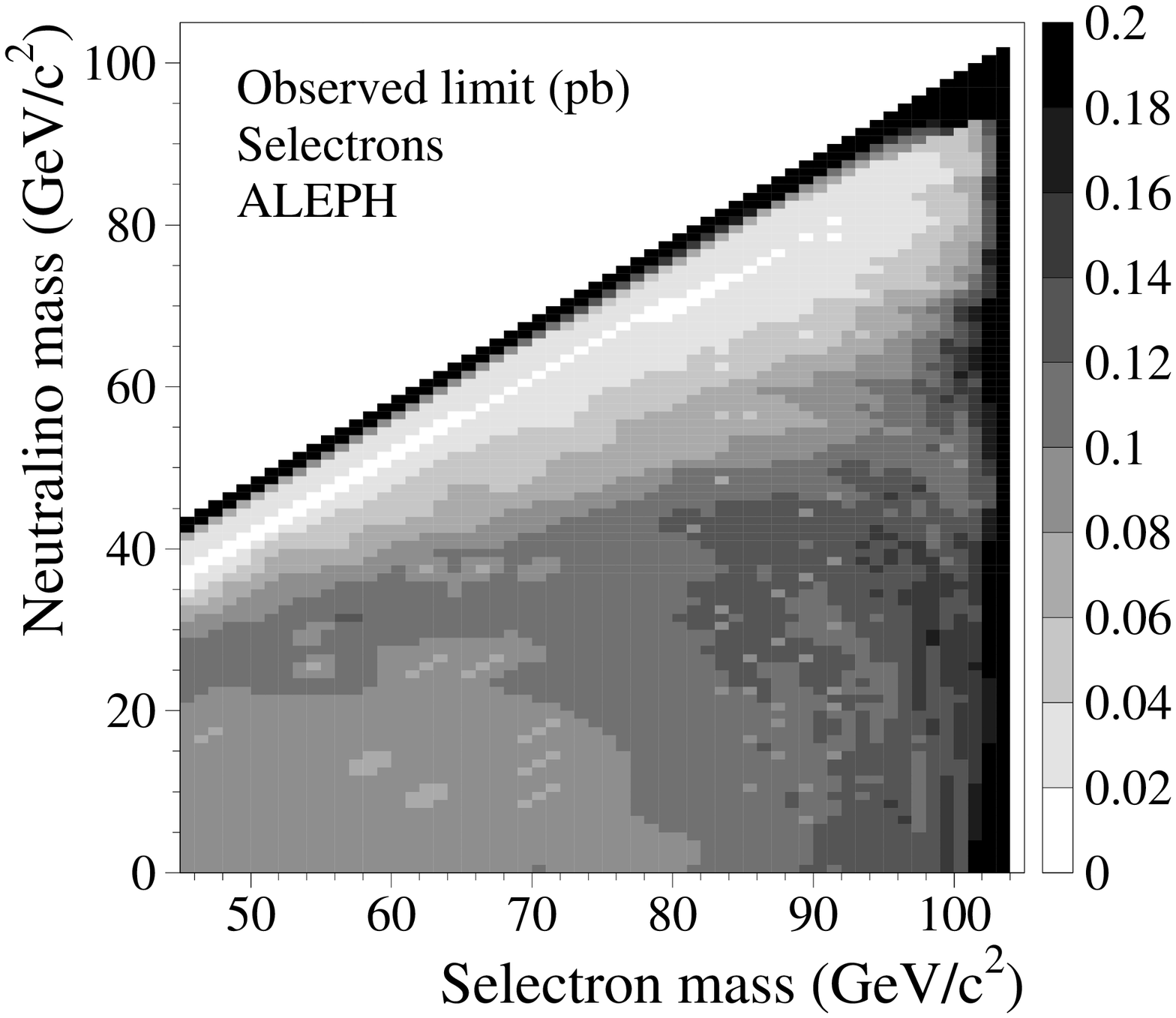,height=7.0cm,width=7.0cm}}
\put(-0.3,7.0){ \epsfig{file=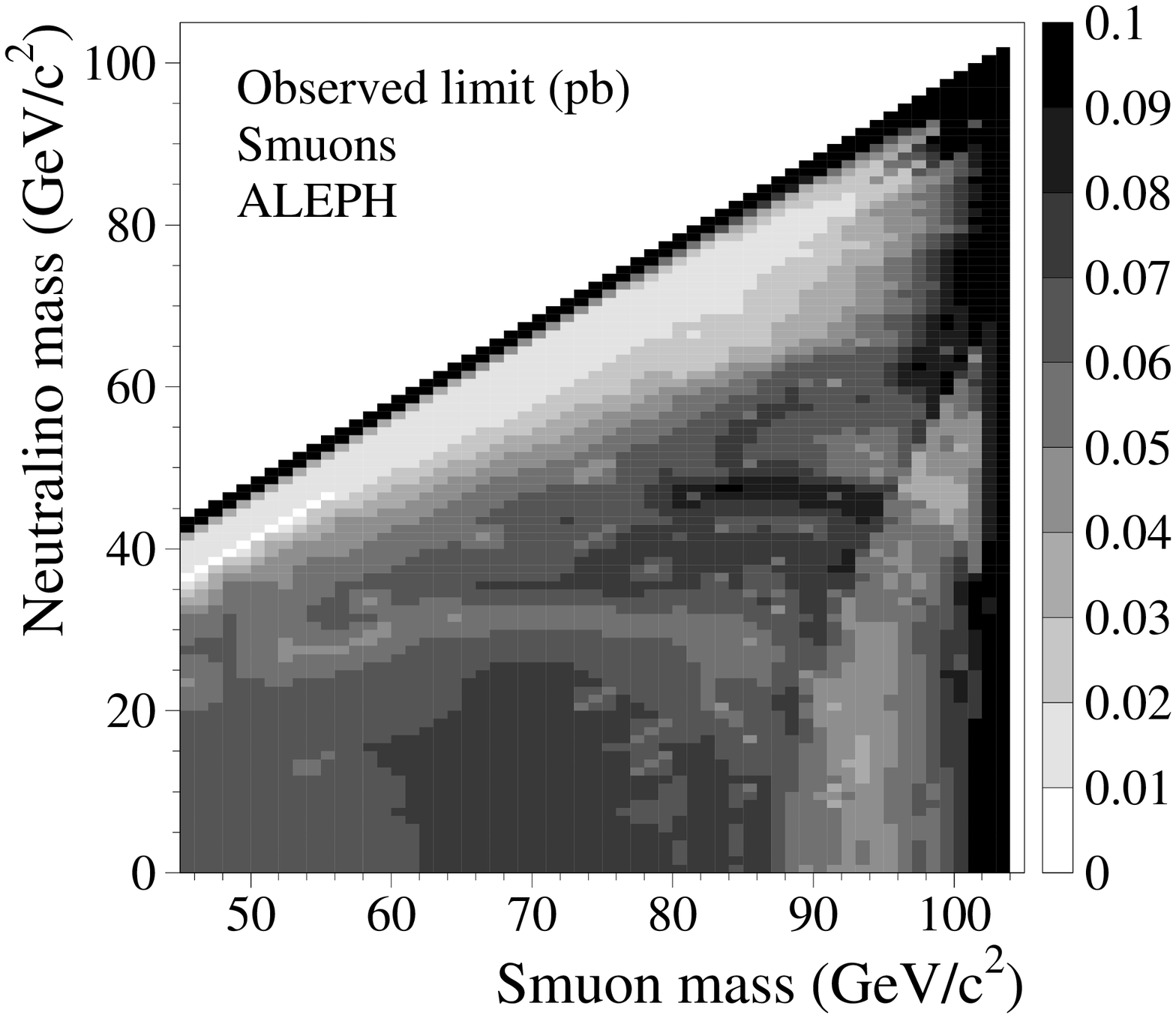,height=7.0cm,width=7.0cm}}
\put(-0.3,0.0){ \epsfig{file=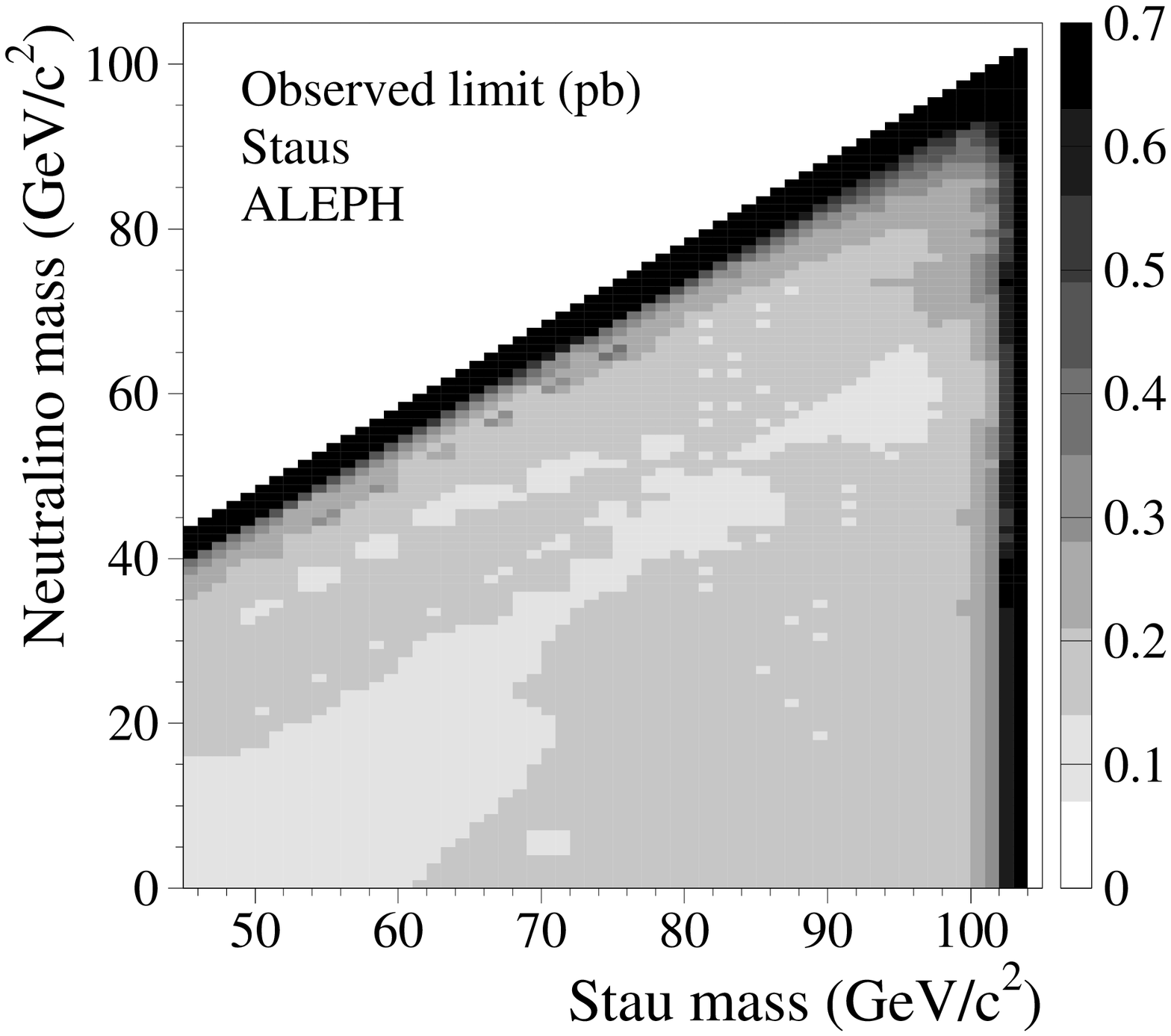,height=7.0cm,width=7.0cm}}
\put(7.5,14.0){ \epsfig{file=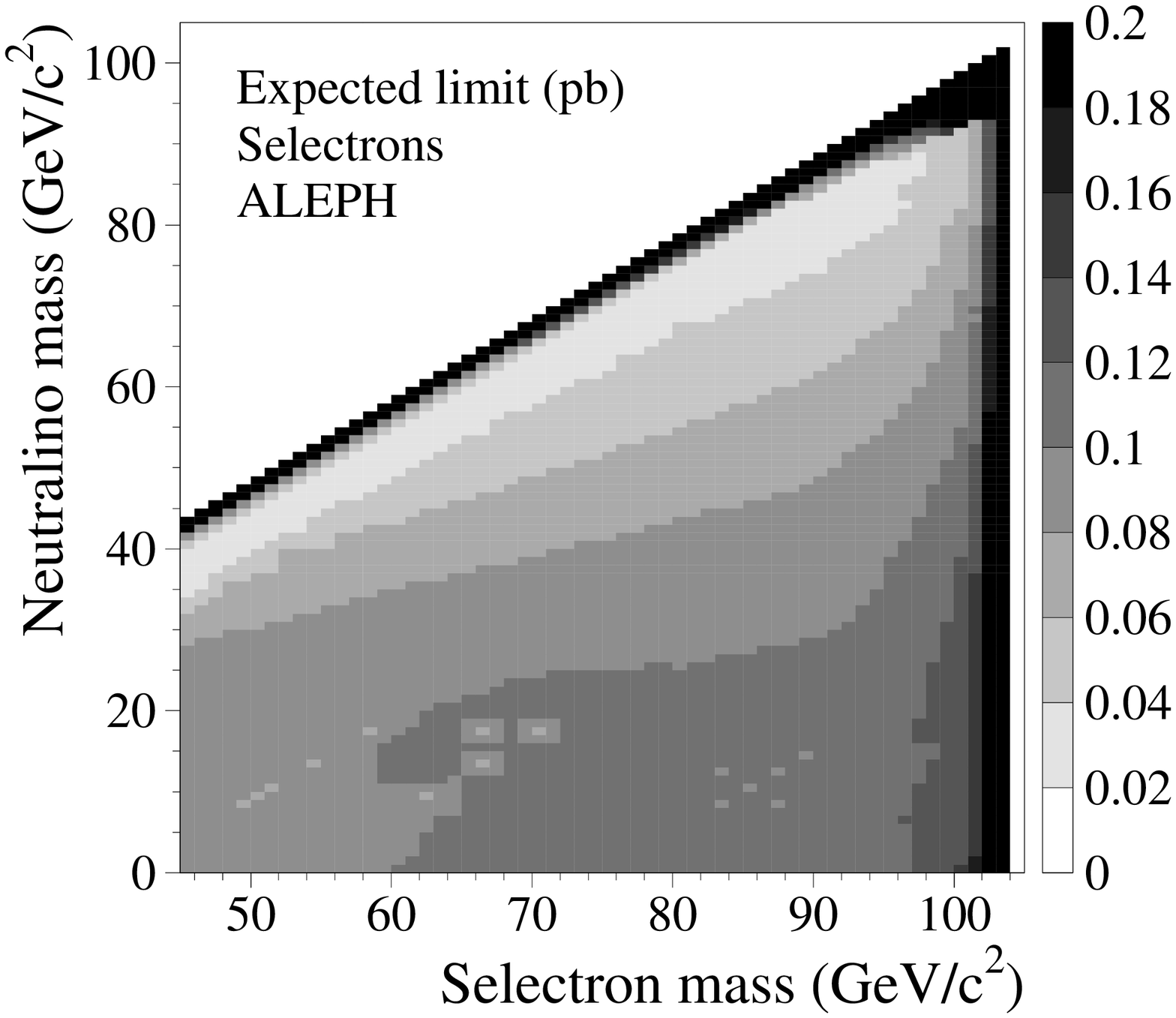,height=7.0cm,width=7.0cm}}
\put(7.5,7.0){  \epsfig{file=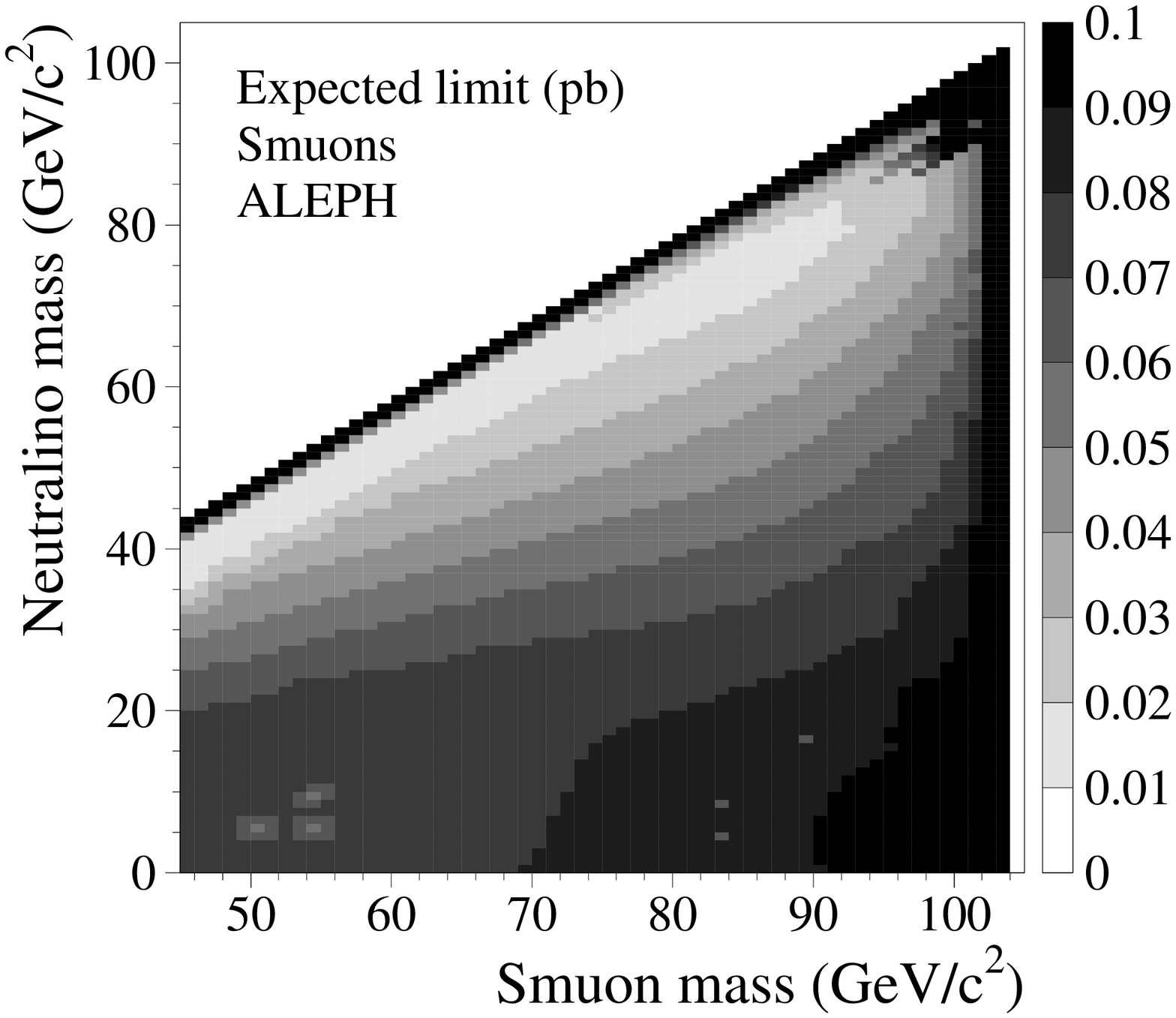,height=7.0cm,width=7.0cm}}
\put(7.5,0.0){  \epsfig{file=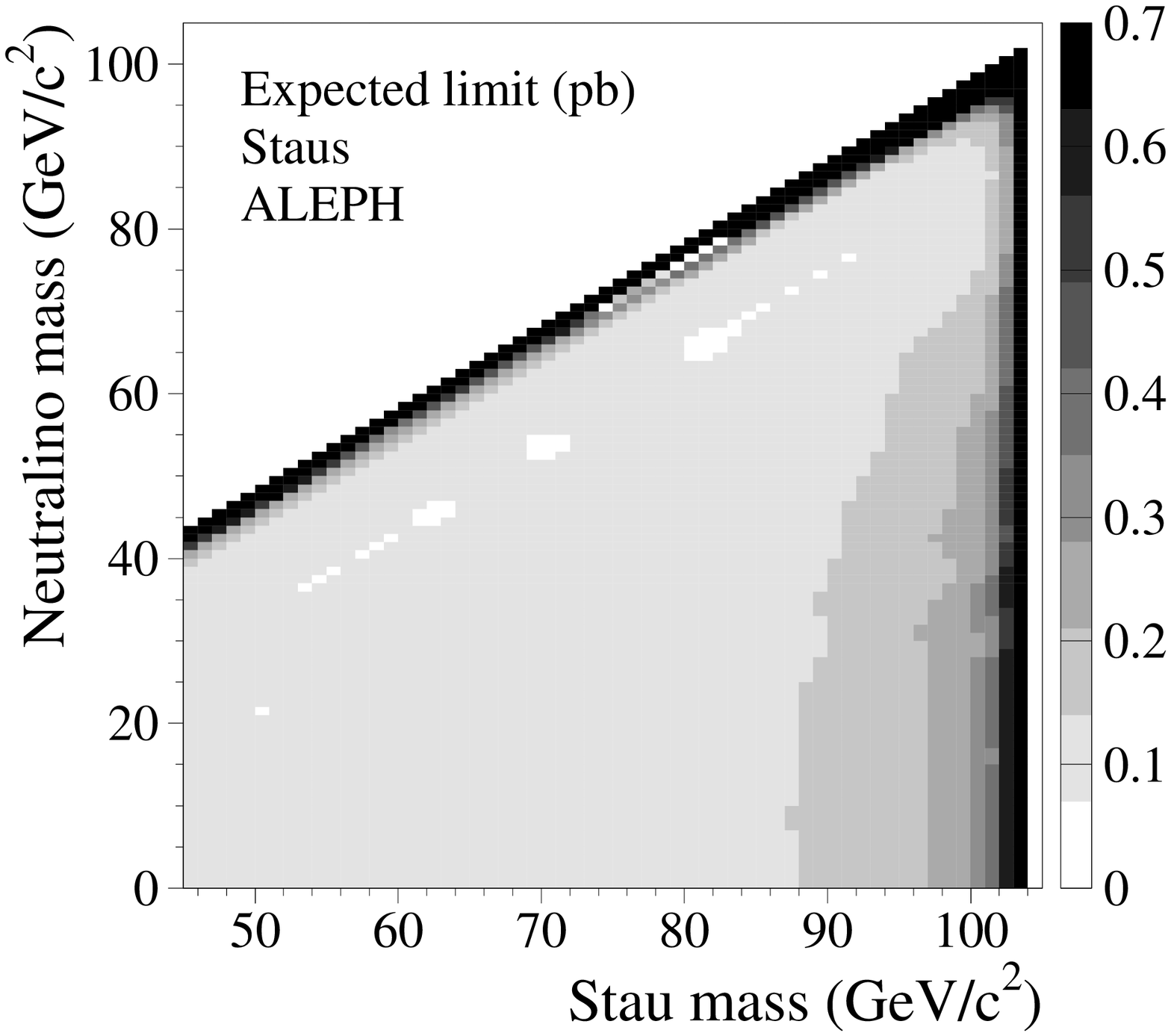,height=7.0cm,width=7.0cm}}
\put(7.0,21.0){(a)}
\put(7.0,14.0){(b)}
\put(7.0,6.9) {(c)}
\end{picture}
\caption{\small Upper limit at 95\% C.L.
   on the observed (left) and expected (right)
   production cross section at $\sqrt{s} = 208$ GeV, times the 
   branching ratio
   into $\ell \Chiz_1$ for (a) selectrons, 
   (b) smuons and (c) staus.}
\label{fig:upper}
\end{center}
\end{figure}

The resulting bounds on the cross section rescaled at 
$\sqrt{s} = 208$\,GeV are shown in Fig.~\ref{fig:upper}.
Cross sections exceeding~0.1\,(0.2)\,pb are excluded in 
a significant part of the plane $(m_{\sLep}, m_{\Chiz_1})$
for selectrons and smuons\,(staus).
In the small-\dm\ region, the limits are less stringent
because the softer decay leptons cause a drop in the selection efficiency.
The limits are also less stringent for slepton masses close to the kinematic 
limit, where only part of the data contributes.

\subsection{Mass exclusion limits}
\label{sec:massl}

In the MSSM framework, the previous bounds allow limits to be set on
the slepton masses as a function of the neutralino mass.
To this end, cross sections and branching ratios are calculated 
with the program {\tt SUSYGEN}~\cite{SUSYGEN98}.
For selectrons and smuons, mixing is expected to be 
negligible and limits are derived under the conservative assumption
that only \sLepR \sLepR\ production contributes. 
Limits are determined for staus in mixed and unmixed scenarios.
Without mixing, conservative
limits are set again under the
assumption that only \sTauR \sTauR\ production contributes.
If the no-mixing assumption is relaxed, the most conservative
limit on the mass of the lightest stau \sTaone\ is obtained with a
mixing angle $\theta_{\sTau}$ which minimizes the production 
cross section, {\it i.e.}, such
that \sTaone\ decouples from the Z boson
($\theta_{\sTau} \simeq 52^{\circ}$). 

The unification of gaugino masses at the GUT scale is assumed
for the computations of the slepton branching ratios 
and of the selectron production cross section.
Branching ratios for the slepton decay are calculated 
for $\mu = -200$\,\gevct\ and $\tanb = 2$.
The branching ratio into $\ell^{\pm} \Chiz_1$
is nearly 100\% except for small neutralino
masses, in which case the cascade decay into $\ell^{\pm} \Chiz_2$ 
followed by $\Chiz_2 \rightarrow \Chiz_1 {\rm f \bar{f}}$ or 
$\Chiz_1 \gamma$ may become kinematically allowed.
The limits are computed under the conservative assumption that
the selection efficiency for decay channels other than 
$\sLep^{\pm} \rightarrow \ell^{\pm} \Chiz_1$
is zero.

\begin{figure}[htbp]
\begin{center}
\setlength{\unitlength}{1.0cm}
\begin{picture}(17.0,17.0)
\put(-0.5,8.5){\epsfig{file=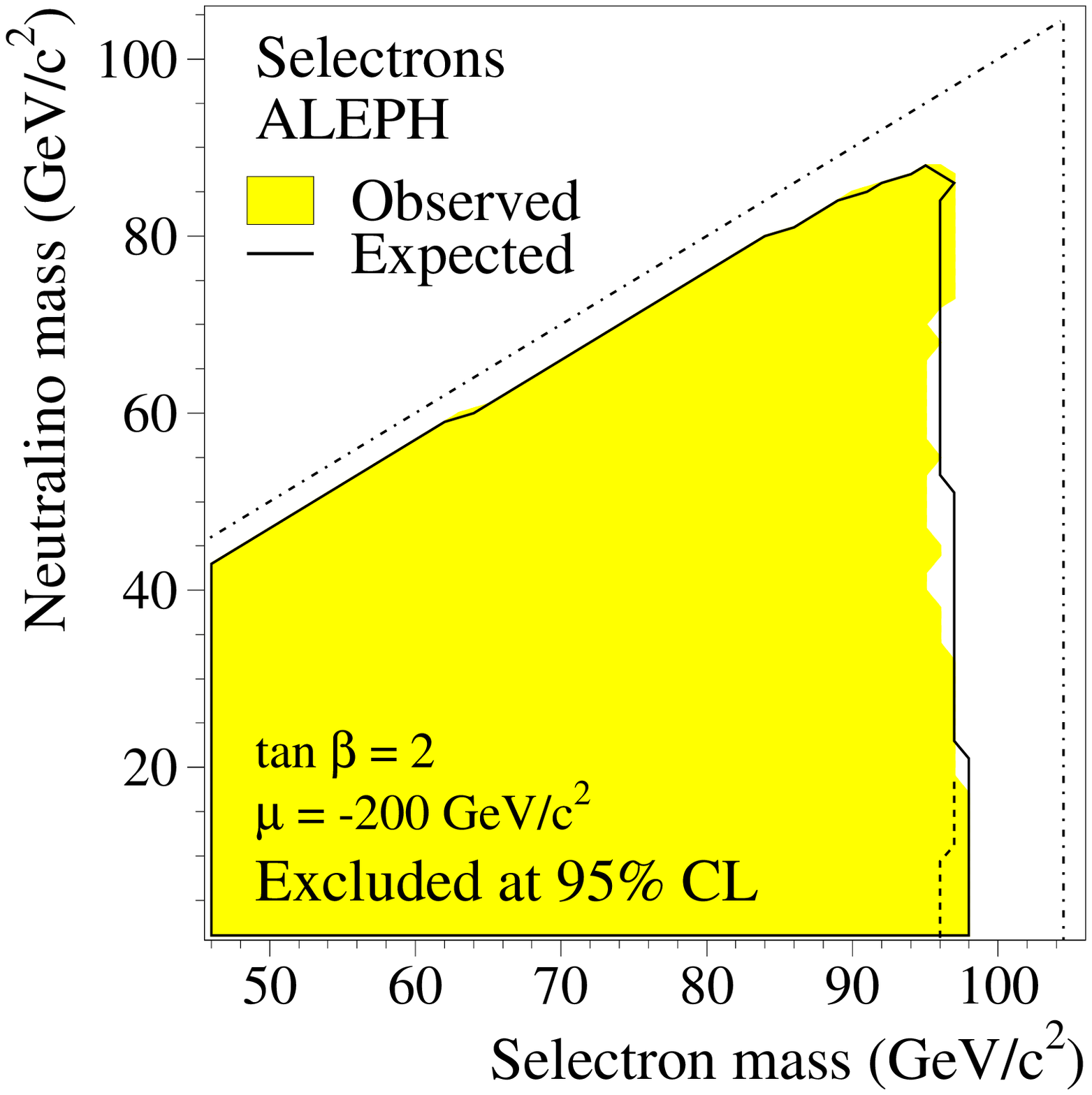,height=9.0cm,width=8.5cm}}
\put(8.0,8.5){ \epsfig{file=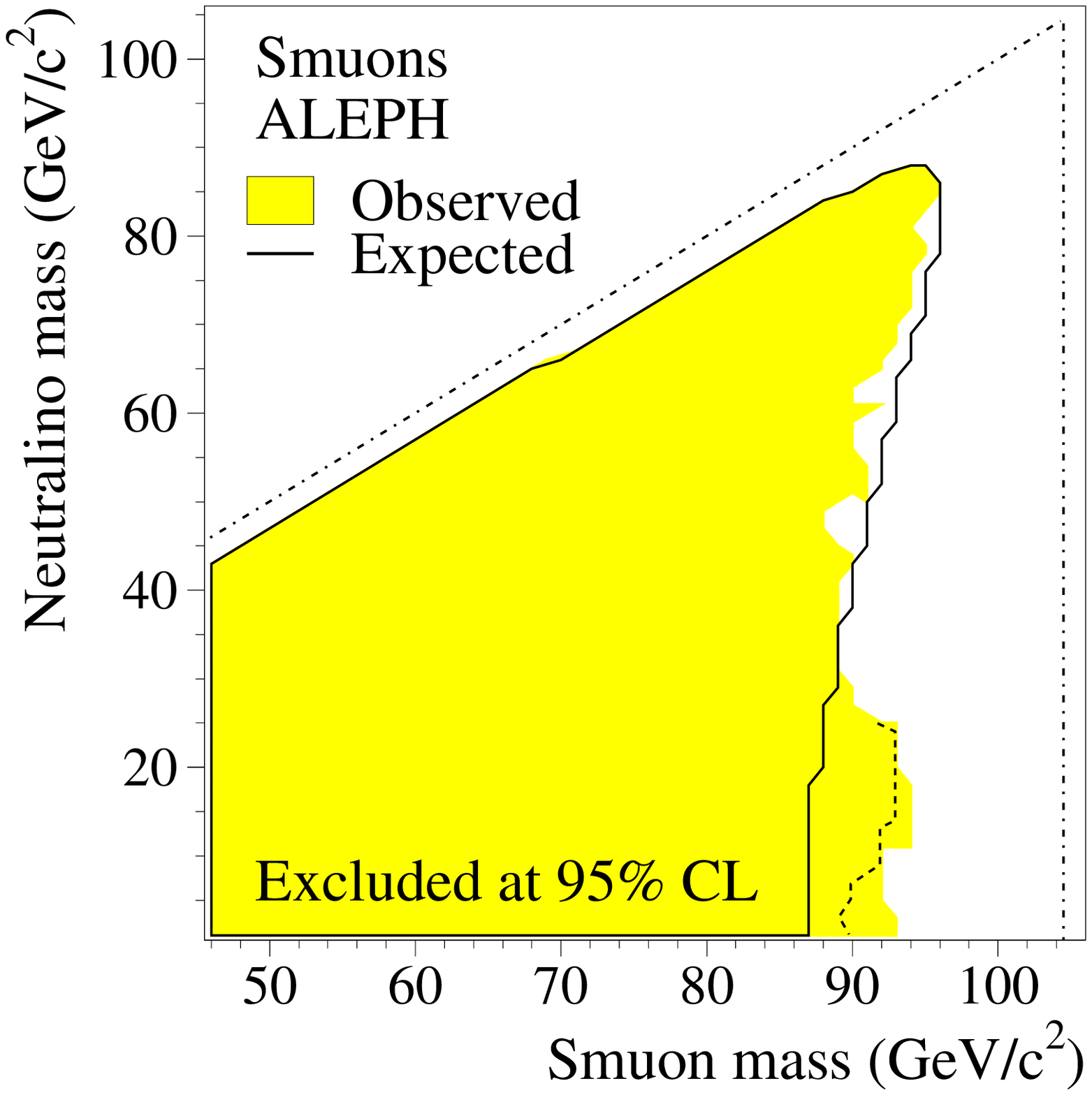,height=9.0cm,width=8.5cm}}
\put(3.8,0.0){ \epsfig{file=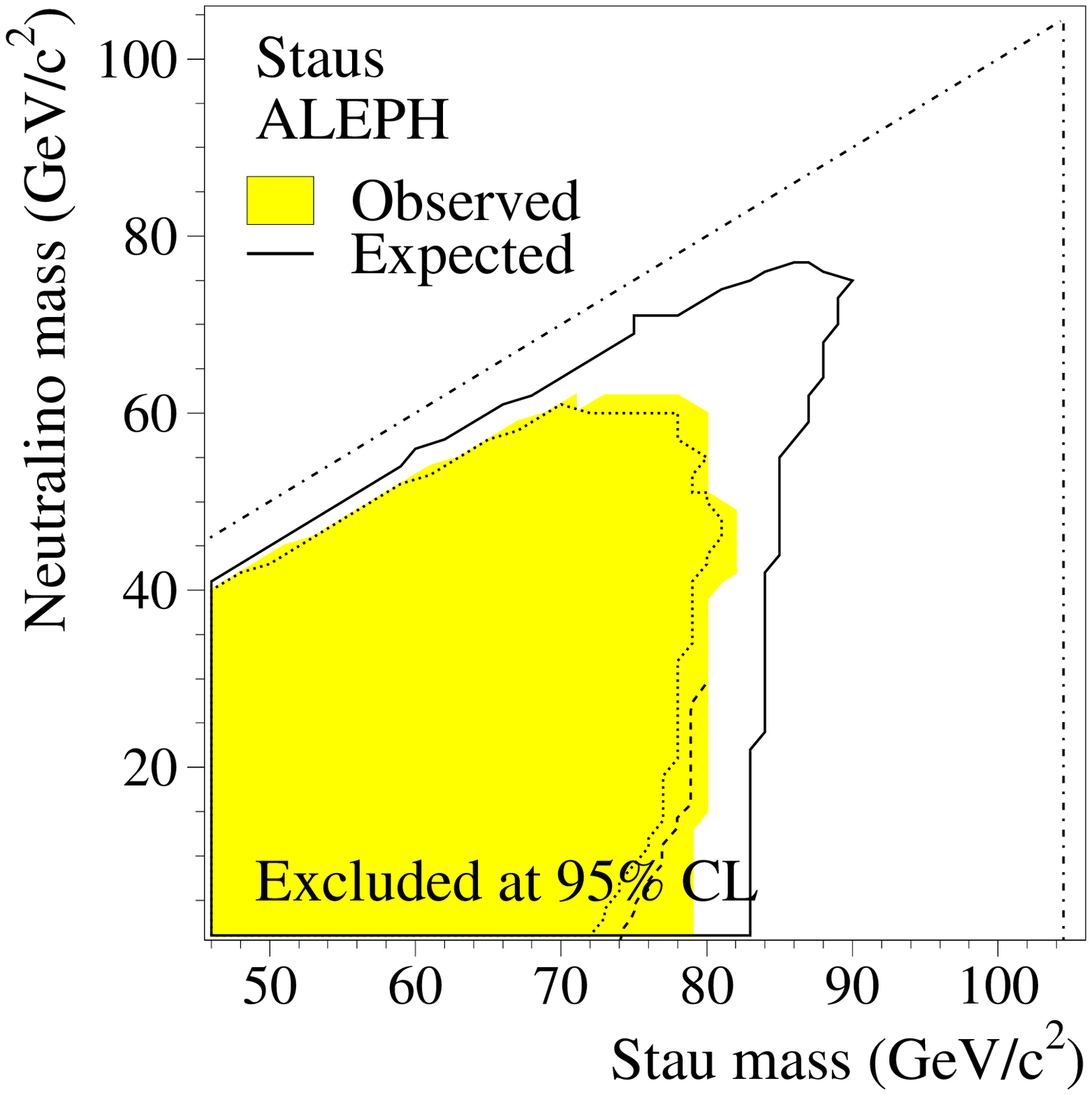,height=9.0cm,width=8.5cm}}
\end{picture}
\caption{\small Excluded regions at 95\% C.L. in the $m_{\sLepR}$ 
    versus $m_{\Chiz_1}$ plane from slepton searches.
    The observed (shaded area) and 
    expected (solid curve) limits are given for 
    BR($\sLep^{\pm} \rightarrow \ell^{\pm} \Chiz_1) = 1$.
    For selectrons, $\mu = -200$\,\gevct\ and $\tanb = 2$
    are assumed.
    The dashed curves give the observed limits when 
    slepton cascade decays are taken into account. 
    For staus, the dotted curve gives the most conservative 
    limit, obtained for 
    $\theta_{\sTau} \simeq 52^{\circ}$ and  
    with cascade decay taken into account.
    For the dashed and dotted curves, 
    BR ($\sLep^{\pm} \rightarrow \ell^{\pm} \Chiz_1$)
    is computed with the values
    $\mu = -200$\,\gevct\ and $\tanb = 2$, and a zero 
    efficiency for the selection of the
    topologies arising from cascade decay is assumed.
    The kinematically accessible region at 209\,GeV is indicated by
    the vertical dash-dotted lines. The diagonal dash-dotted lines
    show the boundary of the domain allowed in the MSSM for a 
    neutralino LSP.}
\label{fig:mass_lim}
\end{center}
\end{figure}

The 95\% C.L. expected and observed bounds on the masses of 
selectrons, smuons and staus as a function of the neutralino 
mass are displayed in Fig.~\ref{fig:mass_lim}, together with
the effect of cascade decays. For staus, the limit obtained 
with mixing angle $\theta_{\sTau} \simeq 52^{\circ}$ is also given.
The observed stau mass limit is less stringent than
expected because a small excess of events was observed 
in 1999~\cite{aleph_slep_1999}.

The 95\% C.L. bounds on the slepton masses for \dm\ $>$ 15\,\gevct\
are summarized in Table~\ref{tab:mass_lim}, with the hypothesis 
BR($\sLep^{\pm} \rightarrow \ell^{\pm} \Chiz_1) = 1$.
For smuons and staus, these limits are independent of the
MSSM parameters. For selectrons, complementary
analyses and a scan on the relevant MSSM parameters 
allow an absolute mass lower limit to be extracted.
The result of such a study is presented in Ref.~\cite{ulla}.

\begin{table}[tb] 
\begin{center} 
\caption{\small Lower limits at 95\% C.L. on the slepton masses for
\dm\ $>$ 15\,\gevct, with the
  hypothesis BR($\sLep^{\pm} \rightarrow \ell^{\pm} \Chiz_1) = 1$.
  For staus, the bound corresponding to the minimum cross section
  (as explained in the text) is also given. In the case of 
  selectrons, the limits are given for 
  $\mu = -200$\,\gevct\ and $\tanb = 2$. \vspace*{0.5cm} }
\vspace*{0.2cm}
\begin{tabular}{|c|c|c|} \hline
 \hline
 \rule{0pt}{4.6mm}
 Channel      & $m(\sLep) >$ & $m(\sLep) >$   \\ 
              & obtained   & expected       \\ \hline
 \sElR        &  95 \gevct    & 96  \gevct        \\
 \sMuR        &  88 \gevct    & 87  \gevct        \\
 \sTauR       &  79 \gevct    & 83  \gevct        \\
 \sTau (min)  &  76 \gevct    & 81  \gevct        \\
  \hline
  \hline
\end{tabular}
\label{tab:mass_lim}
\end{center}
\end{table}

\section{Conclusions}
\label{sec:conc}

Searches for scalar lepton pair production 
have been performed in the data sample
collected by the ALEPH experiment 
at LEP2 at centre-of-mass energies up to 209\,\gev.
The numbers of
candidate events observed are consistent with the background
expected from Standard Model processes. 

In the framework of the MSSM, 95\% C.L. mass exclusion regions have been  
obtained for selectrons, smuons and staus in the plane 
($m_{\sLep}, m_{\Chiz_1}$).
In particular, mass lower limits for 
smuons and staus are set at 88 and 76\,\gevct, respectively,
for a mass difference between the slepton  and the
lightest neutralino in excess of 15\,GeV/$c^2$ and for a slepton decay
branching ratio into $\ell \Chiz_1$ of 100\%.
These limits are independent of the MSSM parameters.
Under the same assumptions, and with $\mu = -200$\,\gevct\ and 
$\tanb = 2$, a lower limit of 95\,GeV/$c^2$ is set 
on the selectron mass.

\section*{Acknowledgements}

It is a pleasure to congratulate our colleagues from the
accelerator divisions for the outstanding operation of LEP\,2, especially in
its last year of running during which the accelerator performance was
pushed beyond expectation. We are indebted to the engineers
and technicians in all our institutions for their contributions to the
excellent performance of ALEPH. Those of us from non-member states wish to
thank CERN for its hospitality and support.



\begin{thebibliography}{99}

\bibitem{mssm} 
  H.P.~Nilles, \PR{110}{1984}{1}; \\
  H.E.~Haber and G.L.~Kane, \PR{117}{1985}{75}.

\bibitem{aleph_slep_1997}
  ALEPH Coll.,
  ``Search for sleptons in \ee\ collisions at centre-of-mass
  energies of 161 and 172\,GeV", \PLB{407}{1997}{377}; \\
  ``Search for sleptons in \ee\ collisions at centre-of-mass
  energies up to 184\,GeV, \PLB{433}{1998}{176}.

\bibitem{aleph_slep_1998}
  ALEPH Coll.,
  ``Searches for sleptons and squarks in \ee\ collisions at 189\,GeV",
  \PLB{469}{1999}{303}.
%
\bibitem{aleph_slep_1999}
  ALEPH Coll., 
  ``Search for supersymmetric particles in \ee\ collisions at $\sqrt{s}$ up 
  to 202\,GeV and mass limit for the lightest neutralino",
  \PLB{499}{2001}{67}. 

\bibitem{lep_slep}
  DELPHI Coll.,
  ``Limits on the masses of supersymmetric particles 
  at $\sqrt{s} = 189$ GeV",
  \PLB{489}{2000}{38}; \\
  L3 Coll., 
  ``Search for scalar leptons in \ee\ collisions at $\sqrt{s} = 189$ GeV",
  \PLB{471}{1999}{280}; \\
  OPAL Coll.,
  ``Search for anomalous production of acoplanar di-lepton events 
  in \ee\ collisions at $\sqrt{s} = 183$ and 189 GeV",
  \EPJ{14}{2000}{51}. 

\bibitem{aleph_det} 
  ALEPH Coll., ``ALEPH: A detector for electron-positron
  annihilations at LEP",
  \NIMA{294}{1990}{121}.

\bibitem{aleph_det2} 
  ALEPH Coll., ``Performance of the ALEPH detector at LEP",
  \NIMA{360}{1995}{481}.

\bibitem{durham}
 Yu.L.~Dokshitzer, J. Phys. {\bf G17} (1991) 1441.

\bibitem{SUSYGEN98} 
  S.~Katsanevas, and P.~Morawitz, ``SUSYGEN 2.2 - A Monte Carlo event 
  generator for MSSM sparticle production at \ee\ colliders",
  \CPC{112}{1998}{227}.

\bibitem{photos}
  E.~Barberio and Z.~W\c{a}s,  \CPC{79}{1994}{291}.

\bibitem{tauola}
  S.~Jadach, Z.~W\c{a}s, R.~Decker and J.H.~K\"{u}hn, 
  \CPC{76}{1993}{361}.

\bibitem{bhwide}
  S.~Jadach, W.~Placzek and B.F.L.~Ward,
 \PLB{390}{1997}{298}.

\bibitem{koralz}
  S.~Jadach and Z.~W\c{a}s, \CPC{36}{1985}{191}.

\bibitem{phot02}
  J.A.M.~Vermaseren, in ``Proceedings of the IVth International 
  Workshop on Gamma Gamma interactions", Eds. G. Cochard and P. Kessler,
  Springer Verlag, 1980.

\bibitem{koralw}
  M.~Skrzypek, S.~Jadach, W.~Placzek and Z.~W\c{a}s, 
  \CPC{94}{1996}{216}.

\bibitem{pythia}
  T.~Sj\"{o}strand, \CPC{82}{1994}{74}.

\bibitem{geant}
  ``GEANT Detector description and simulation tool", CERN Program Library, 
  CERN-W5013 (1993).

\bibitem{nbar95} 
  J.F.~Grivaz and F.~Le Diberder, ``Complementary analysis and 
  acceptance optimization in new particle searches", 
  LAL preprint \# 92-37 (1992); \\
  ALEPH Coll., ``Mass limit for the standard model higgs boson 
  with the full LEP 1 ALEPH data sample", \PLB{384}{1996}{427}.

\bibitem{fisher}
  R.A.~Fisher, ``The use of multiple measurements in taxonomic problems",
  Annals of Eugenics {\bf 7} (1936) 179.

\bibitem{Read} 
  A.L.~Read, ``Modified frequentist analysis of search results
  (The CLs Method)", CERN-OPEN-2000-205, 81.

%
\bibitem{Cousins} 
  R.D.~Cousins and V.L.~Highland,
  \NIMA{320}{1992}{331}.

%
\bibitem{ulla} 
ALEPH Coll., ``Lower mass limit for 
the MSSM selectron and sneutrino
obtained with the ALEPH detector", contribution~\#~236 to the 
IECHEP, Budapest, Hungary, 12-18 July 2001, ALEPH-CONF 2001-047;\\
http://alephwww.cern.ch/ALPUB/oldconf/summer01/33/sel\_limit.ps. \\

\end{thebibliography}
\end{document}